\documentclass[useAMS,usenatbib]{mn2e}
\usepackage{natbib}
\usepackage{graphicx}
\usepackage{amssymb}
\usepackage{amsmath}
\usepackage{calc}

\title[The non-symmetric ion-atom processes]
{The non-symmetric ion-atom radiative processes in the stellar atmospheres}

\author[A. A. Mihajlov et al]{A. A. Mihajlov$^{1,3}$\thanks{E-mail: mihajlov@ipb.ac.rs}, Lj. M. Ignjatovi{\' c}$^{1,3}$, V. A. Sre{\'
c}kovi{\' c}$^{1,3}$,
\newauthor M. S. Dimitrijevi{\'c}$^{2,3}$ and
A. Metropoulos $^{4}$\\
$^{1}$University of Belgrade,Institute of Physics, P. O. Box 57, 11001 Belgrade,
Serbia\\
$^{2}$Astronomical Observatory, Volgina 7, 11160 Belgrade 74,
Serbia\\
$^{3}$Isaac Newton Institute of Chile, Yugoslavia Branch\\
$^{4}$Theoretical and Physical Chemistry Institute, NHRF, Athens,
Greece}

\begin{document}

\date{}

\pagerange{\pageref{firstpage}--\pageref{lastpage}} \pubyear{2008}

\maketitle

\label{firstpage}

\begin{abstract}

The aim of this research is to show that the processes of absorption
charge-exchange and photo-association in $A+B^{+}$ collisions together with
the processes of $AB^{+}$ photo-dissociation in the case of strongly
non-symmetric ion-atom systems, significantly influence the opacity of
stellar atmospheres in ultraviolet (UV) and extreme UV (EUV) region. In this work, the significance of such
processes for solar atmosphere is studied. In the case of the solar atmosphere
the absorption processes with $A=$ H and $B=$ Mg and Si are treated as
dominant ones, but the cases $A=$ H and $B=$ Al and $A=$ He and $B=$ H are also
taken into consideration. The choice of just these species is caused by
the fact that, of the species relevant for the used solar-atmosphere model,
it was only for them that we could determine the necessary characteristics of the
corresponding molecular ions, i.e. the molecular potential curves and dipole
matrix elements. It is shown that the efficiency of the examined non-symmetric
processes within the rather wide corresponding quasi-molecular absorption bands
in the far-UV and EUV regions is comparable and sometimes even greater than the
intensity of the known symmetric ion-atom absorption processes, which are
included now in the models of the solar atmosphere. Consequently, the presented
results suggest that the non-symmetric ion-atom absorption processes also have
to be included \emph{ab initio} in the corresponding models of the stellar
atmospheres.
\end{abstract}


\section{Introduction}

Significant influence of at least some of the ion-atom radiative processes
on the optical characteristics of the solar atmosphere has already been
established. Here we mean the following symmetric processes of
molecular ion photo-dissociation/association and radiative charge exchange
in ion-atom collisions
%%%%%%%%%%%%%%%%%%%%%%%%%%%%%%%%%%%%%%%%%%%%%%%%%%%%%%%%%%%%%%%%%%%%
\begin{equation}
\label{eq:sim1} \varepsilon_{\lambda}
+ \textrm{H}_{2}^{+} \longleftrightarrow \textrm{H} + \textrm{H}^{+},
\end{equation}
%%%%%%%%%%%%%%%%%%%%%%%%%%%%%%%%%%%%%%%%%%%%%%%%%%%%%%%%%%%%%%%%%%%%
%%%%%%%%%%%%%%%%%%%%%%%%%%%%%%%%%%%%%%%%%%%%%%%%%%%%%%%%%%%%%%%%%%%%
\begin{equation}
\label{eq:sim2} \varepsilon_{\lambda}
+ \textrm{H}^{+} + \textrm{H} \longleftrightarrow \textrm{H} + \textrm{H}^{+},
\end{equation}
%%%%%%%%%%%%%%%%%%%%%%%%%%%%%%%%%%%%%%%%%%%%%%%%%%%%%%%%%%%%%%%%%%%%
which were studied in the context of the atmosphere of the Sun in \citep{mih86,
mih93a, mih94b, mih07a}. Let us note that here H = H(1s), H$_{2}^{+}$ is the
molecular ion in the ground electronic state, and $\varepsilon_{\lambda}$ - the
energy of a photon with wavelength $\lambda$. Of course, the results obtained
in the mentioned articles are significant and for atmospheres of other solar or near solar type stars.

\begin{figure}
\begin{center}
\includegraphics[width=\columnwidth,
height=0.75\columnwidth]{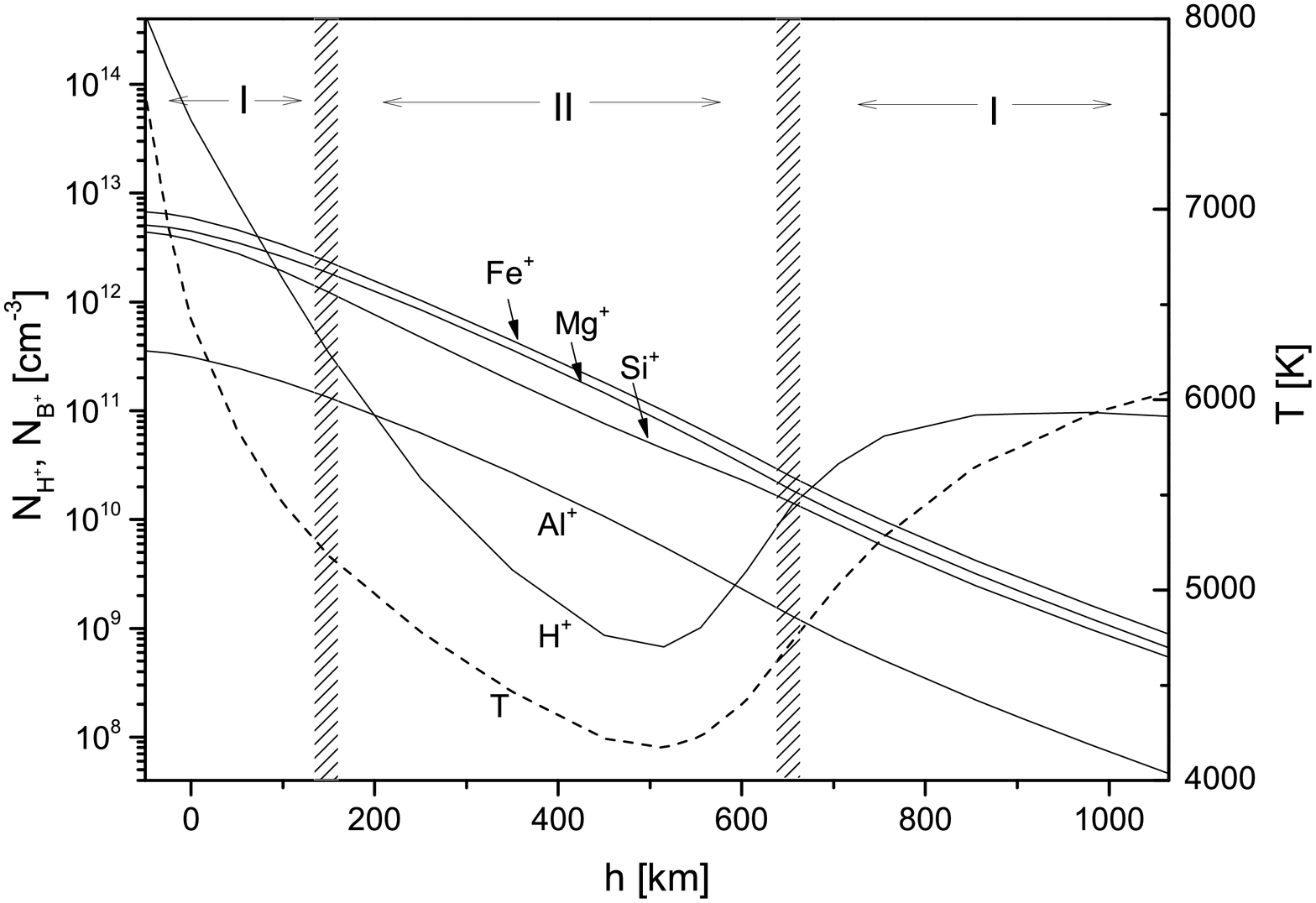} \caption{The behavior of the temperature $T$ and the
densities $N_{\textrm{H}^{+}}$ and $N_{B^{+}}$ of the ions H$^{+}$ and the metal ions $B^{+}$ for
the non-LTE model C from \citet{ver81} within the solar atmosphere.}
\label{fig:abund}
\end{center}
\end{figure}

Only the processes (\ref{eq:sim1}) and (\ref{eq:sim2}) were taken into account in
the mentioned papers, since the contribution of other symmetric ion-atom radiative
processes to the solar-atmosphere opacity could be completly neglected due to
the composition of the atmosphere, while the possible non-symmetric processes
were excluded from the consideration because of the orientation of the research,
alredy established in the first paper \citep{mih86}, towards the visible and near
UV and IR parts of the electro-magnetic (EM) spectrum. However, in \citet{mih07a}
it was demonstrated that the efficiency of the processes (\ref{eq:sim1}) and
(\ref{eq:sim2}) becomes close to the total efficiency of the concurrent
electron-ion and electron-atom radiative processes outside of these parts
of the EM spectrum, namely in far UV and EUV regions. It is important that just
these spectral regions are very significant in the case of the solar atmosphere.
This is caused by the fact that the solar emission in far UV and EUV regions very
strongly affects the ionosphere every day, and by extension the whole of the
Earth's atmosphere. Therefore the solar EM emission in the mentioned
regions has been the object of extensive investigation for a long time (see the
classic book: \cite{whi77}), which continues up until now (see e.g. \citet{wor01,
woo08, woo09}). It is clear that in this context it becomes necessary to pay
attention not only to the symmetrical ion-atom processes (\ref{eq:sim1}) and
(\ref{eq:sim2}), but also to each new process which might affect the
mechanisms of EM radiation transfer in far UV and EUV regions in the solar
atmosphere, and consequently the corresponding optical characteristics. These
facts suggested that it could be useful to carefully examine also the possible
influence of the relevant non-symmetric ion-atom radiative processes on the
solar-atmosphere opacity, namely
%%%%%%%%%%%%%%%%%%%%%%%%%%%%%%%%%%%%%%%%%%%%%%%%%%%%%%%%%%%%%%%%%%%%
\begin{equation}
\label{eq:nonsim1} \varepsilon_{\lambda}
+ AB^{+} \longrightarrow A^{+} + B,
\end{equation}
%%%%%%%%%%%%%%%%%%%%%%%%%%%%%%%%%%%%%%%%%%%%%%%%%%%%%%%%%%%%%%%%%%%%
\begin{equation}
\label{eq:nonsim2} \varepsilon_{\lambda}
+ A + B^{+} \longrightarrow A^{+} + B,
\end{equation}
%%%%%%%%%%%%%%%%%%%%%%%%%%%%%%%%%%%%%%%%%%%%%%%%%%%%%%%%%%%%%%%%%%%%
\begin{equation}
\label{eq:nonsim3} \varepsilon_{\lambda}
+ A + B^{+} \longrightarrow (AB^{+})^{*} ,
\end{equation}
%%%%%%%%%%%%%%%%%%%%%%%%%%%%%%%%%%%%%%%%%%%%%%%%%%%%%%%%%%%%%%%%%%%%
where $B$ is an atom in the ground state with its ionization potential $I_{B}$
smaller than the ionization potential $I_{A}$ of the atom $A$, while $AB^{+}$ and
$(AB^{+})^{*}$ are the corresponding molecular ions in the electronic states
which are asymptotically correlated with the states of the systems $A + B^{+}$
and $A^{+} + B$ respectively, and the possible partners are determined by the
used solar-atmosphere models. One can see that the processes (\ref{eq:nonsim1})
and (\ref{eq:nonsim2}) represent the analogues of the processes (\ref{eq:sim1})
and (\ref{eq:sim2}), while the process (\ref{eq:nonsim3}) does not have
a symmetric analogue. In this work the standard non-LTE model C for the solar
atmosphere from \citet{ver81} is used. The reason is the fact that as yet all
the relevant data needed for our calculations are provided in the tabular form
only for this model, and that in \citet{sti02} the solar atmosphere
model C from \cite{ver81} is treated as an adequate non-LTE model. In accordance
to the chosen model here we take into account the non-symmetric processes
(\ref{eq:nonsim1}-\ref{eq:nonsim3}) with $A=$ H(1s) and $B=$ Mg, Si, Fe and
Al, as well as with $A=$ He(1s$^2$) and $B=$ H(1s).
%%%%%%%%%%%%%%%%%
For the solar photosphere the behavior of the densities of the metal Mg$^{+}$,
 Si$^{+}$, Fe$^{+}$ and Al$^{+}$ ions is particulary important. Namely, in
accordance with the Tab's 12, 17 and 19-22 from \cite{ver81} this behavior, as
well as the behavior of the temperature $T$ and the ion H$^{+}$ density, within
the solar photosphere can be illustrated by the Fig.\ref{fig:abund}, where $h$ is
the height of the considered layer with the respect to the chosen referent one.
The region of $h$ is chosen here in accordance with the Fig.4 from the previous
paper \cite{mih07a}, where the relative efficiencies of the symmetric ion-atom
processes (\ref{eq:sim1}) and (\ref{eq:sim2}) with  the respect to the relevant
concurrent (electron-atom and electron-ion) radiative processes are presented.
This region is slitted in three parts: two denoted with I, which corresponds to
the areas where the efficiency of the processes (\ref{eq:sim1}) and
(\ref{eq:sim2}) with $A=$ H is close to the one of the mentioned concurrent
processes, and one denoted with II, where their efficiencies can be neglected.
From the figure \ref{fig:abund} one can see that in the parts I the ion H$^{+}$
density dominates with the respect to all metal ion $B^{+}$ densities, which
means that within these parts it is expected that the efficiency of the
symmetric processes (\ref{eq:sim1}) and (\ref{eq:sim2}) is more grater than the
one of the non-symmetric processes (\ref{eq:nonsim1}-\ref{eq:nonsim3}). However,
from the same figure one can see also that:\\
- in the part II, i.e. in the neighborhood of the temperature minimum, each of
the ion $B^{+}$ densities is greater then the ion H$^{+}$ density,\\
- the width of the part II is close to the total width of the parts denoted with
I.
\\
From here it follows that, in the principle, the contribution of the
non-symmetric processes (\ref{eq:nonsim1}-\ref{eq:nonsim3}) to the solar
atmosphere opacity can be comparable to the one of the symmetric processes
(\ref{eq:sim1}) and (\ref{eq:sim2}), since in the both non-symmetric and
symmetric cases as the neutral partner the same atom $\textrm{H}$ appears, and that
symmetric and non-symmetric ion-atom radiative processes together could be
treated as a serious partner to the above mentioned concurrent processes in the
whole solar photosphere.

In the general case of the partially ionized gaseous plasmas, apart of the
absorption processes (\ref{eq:nonsim1}-\ref{eq:nonsim3}), it is necessary to
consider also the corresponding inverse emission processes, namely: the emission
charge exchange and photo-association in $A^{+} + B$ collisions, and the
photo-dissociation of the molecular ion $(AB^{+})^{*}$. However, here only
the absorption processes (\ref{eq:nonsim1}) and (\ref{eq:nonsim2}) have to be
taken into account. Namely, under the conditions from \citet{ver81} the influence
of the emission processes in $A^{+} + B$ collisions on the optical
characteristics of the considered atmospheres can be neglected in comparison to
the other relevant emission processes, since both $A^{+}$ and $B$ partners belong
to the poorly represented components.

From the beginning of our investigations of ion-atom radiative processes
as the final aim we have always the inclusion of the considered processes
in the stellar atmospheres models. Here it should be noted that we considered as
our task only to provide the relevant data (the corresponding spectral absorption
coefficients etc.), which are needed for the stellar atmospheres modeling,
without involving into the process of the modeling itself. Consequently, for us
it was important only to know whether the processes, which we studied in
connection with the considered stellar atmosphere, are included in the
corresponding models or not. So during the previous investigations, whose
results were published in \citet{mih86,mih93a,mih94b,mih07a}, we reliably knew
that, for example, the processes of the radiative charge-exchange \ref{eq:sim2}
generally were not taken into account in connection with the solar atmosphere.
Apart of that, it was known that the photo-dissociation processes \ref{eq:sim1}
were seriously treated only when the atom and ion ($\textrm{H}$ and H$^{+}$) densities
are close, while our results suggested that the processes (\ref{eq:sim1}) and
(\ref{eq:sim2}) are of the greatest importance for the weakly ionized stellar
layers (ion density/atom density $\lesssim 10^{-3}$). These reasons fully
justified the mentioned investigations. It is important
that the situation about the symmetric processes (\ref{eq:sim1}) and
(\ref{eq:sim2}) begins to change now in the positive way, since these
processes are already included in some solar atmosphere models \citep{fon09}.
%%%%% Abund %%%%%%
%%%%%%%%%%%%%%%%%%%%%%%%%%%%

The main aim of this work is to draw attention to the non-symmetric radiative
processes (\ref{eq:nonsim1})-(\ref{eq:nonsim3}) as the factors of the influence
on the solar atmospher opacity in the significant parts of UV and EUV regions
and, in accordance with above mentioned, to show that these processes should be
included \emph{ab initio} in the solar atmosphere models, as well as in the models
of solar and near solar type stars, together with the
symmetric processes (\ref{eq:sim1}) and (\ref{eq:sim2}). In this context we will
have to determine here the corresponding spectral absorption coefficients, as the
functions of $\lambda$, the local temperature $T$ and the relevant particle
densities, for the conditions which correspond to the photosphere of the Sun.
For that purpose the needed characteristics of the considered ion-atom systems, i.e. the
molecular potential curves and dipole matrix elements, are presented in Section 2.
Then, the relevant characteristics of
these processes themselves, i.e. the mean thermal cross-sections for the
photo-dissociation processes (\ref{eq:nonsim1}), and the spectral rate
coefficients for the absorption charge exchange processes (\ref{eq:nonsim2}) and (\ref{eq:nonsim3}),
will be presented in Section 3. With the help of these characteristics in
Section 4 will be calculated the total spectral absorption coefficients,
characterizing (\ref{eq:nonsim1}), (\ref{eq:nonsim2}) and (\ref{eq:nonsim3}) absorption
processes as the functions of $\lambda$ and the position within the solar
photosphere. Finally, the values of the parameters which characterize the
relative contribution of the non-symmetric processes
(\ref{eq:nonsim1})-(\ref{eq:nonsim3}) with the respect to the total contribution
of the symmetric and non-symmetric radiative processes
(\ref{eq:sim1})-(\ref{eq:nonsim3}), which are also calculated in Section 4, presents
the one of the main results of this work. Because of the properties of the
considered strongly non-symmetric ion-atom systems only the far-UV and EUV
regions of $\lambda$ are treated here. Let us note that we were able to determine
the relevant characteristics of the molecular ions $\textrm{H}B^{+}$ for the cases
$B=\textrm{Mg, Si}$ and $\textrm{Al}$, and consequently only these cases are considered within this
work.
%%%%% shema %%%%%%
\begin{figure}
\begin{center}
\includegraphics[width=0.70\columnwidth,
height=0.60\columnwidth]{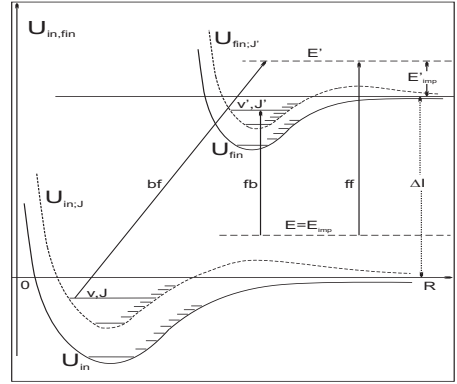} \caption{Schematic presentation of the non-symmetric
processes (\ref{eq:nonsim1}-\ref{eq:nonsim3}) caused by the bf-, ff-, and fb-radiative
transitions: $\Delta I= I_{A}-I_{B}$, where $I_{A}$ and $I_{B}$ are ionization potentials of
the atoms $A$ and $B$; $E=E_{imp}$ and $E'_{impa}$ - the impact energies of the corresponding
ion-atom systems; $U_{in;J}(R)$ and $U_{fin;J'}(R)$ - are the effective potentials, given
by equation (\ref{eq:Ui}).}
\label{fig:HeH+}
\end{center}
\end{figure}
%%%%%%%%%%%%%%%%%%%%%%%%%%%%
\section{The properties of the non-symmetric ion-atom systems}

As in the previous papers the ion-atom radiative processes are described here within two basic
approximations: the adiabatic approximation for the relative motion of the nucleus of the
considered ion-atom systems, and the dipole approximation for the interaction of these systems
with the free electromagnetic field. Since these approximations are discussed in details in
the literature (see for example \cite{mih81a}), the corresponding matter is considered here
briefly, with references only to the elements specific just for the non-symmetric processes
(\ref{eq:nonsim1}-\ref{eq:nonsim3}), which are schematically shown in
Fig. \ref{fig:HeH+}.

In accordance with Fig. \ref{fig:HeH+} the photo-dissociation (bound-free)
processes (\ref{eq:nonsim1}), charge-exchange absorption (free-free) processes
(\ref{eq:nonsim2}), and photo-association (free-bound) processes
(\ref{eq:nonsim3}) are caused by the radiative transitions
%%%%%%%%%%%%%%%%%%%%%%%%%%%%%%%%%%%%%%%%%%%%%%%%%%%%%%%%%%%%%%%%%%%%
\begin{equation}
\label{eq:sheme}
\begin{array}{lll}
              |in;R> |in,J,v;R> \rightarrow |fin;R> |fin,J',E';R>, \\
              |in;R> |in,J,E;R> \rightarrow |fin;R> |fin,J',E';R>,\\
              \displaystyle{|in;R> |in,J,E;R> \rightarrow |fin;R> |fin,J',v';R>, }
             \end{array}
\end{equation}
%%%%%%%%%%%%%%%%%%%%%%%%%%%%%%%%%%%%%%%%%%%%%%%%%%%%%%%%%%%%%%%%%%%%
where $|in;R> |in,J,v;R>$ and $|in;R> |in,J,E;R>$ are initial, and $|fin;R> |fin,J',E';R>$,
$|fin;R> |fin,J',E';R>$ and $|fin;R> |fin,J',v';R>$ are final states of considered
ion-atom system, determined as the products of the adiabatic electronic states $|in;R>$
and $|fin;R>$ and the corresponding states which describe relative nucleus motion, and $R$
denotes the internuclear distance. It is assumed that these transitions are allowed by the
dipole selective rules.

\noindent From Fig.\ref{fig:HeH+} one can see that:\\
- $|in;R>$ and $|fin;R>$ belongs to the groups (i) and (ii) of the states of the molecular
ions $AB^{+}$ and $(AB^{+})^{*}$ which are asymptotically correlated with the electronic
states of the ion-atom systems $A + B^{+}$ and $A^{+} + B$ respectively;\\
-$|in,J,v;R>$ and $|fin,J',v';R>$ are the bound ro-vibrational states of the same molecular
ion, defined by the orbital quantum numbers $J$ and $J'$ and vibrational quantum numbers $v$
and $v'$;\\
-$|in,J,E;R>$ and $|fin,J',E';R>$ are the free states of the same molecular ion defined
by the orbital quantum numbers $J$ and $J'$ and the total energies $E$ and $E'$. Let us
note that as zero of energy here is taken the total energy of the immobile atom $A$ and
ion $B^{+}$ at $R=\infty$.

The states $|in,J,v;R>$, $|in,J,E;R>$, $|fin,J',E';R>$ and $|fin,J',v';R>$ are determined
as the solutions of the corresponding Schrodinger equations
%%%%%%%%%%%%%%%%%%%%%%%%%%%%%%%%%%%%%%%%%%%%%%%%%%%%%%%%%%%%%%
\begin{equation}
\label{eq:b} [ -\frac{1}{2\mu}\Delta + U_{in;J}(R)]
|in,J,v;R \rangle = \epsilon_{J,v} \cdot |in, J, v;R \rangle ,
\end{equation}
%%%%%%%%%%%%%%%%%%%%%%%%%%%%%%%%%%%%%%%%%%%%%%%%%%%%%%%%%%%%%%
\begin{equation}
\label{eq:ff} [ -\frac{1}{2\mu}\Delta + U_{in;J}(R)]
|in,J,E;R\rangle = E\cdot |in,J,E; R\rangle ,
\end{equation}
%%%%%%%%%%%%%%%%%%%%%%%%%%%%%%%%%%%%%%%%%%%%%%%%%%%%%%%%%%%%%%
\begin{equation}
\label{eq:ff}
[-\frac{1}{2\mu}\Delta + U_{fin;J'}(R)]
|fin,J',E';R\rangle = E'_{imp}\cdot |fin,J',E';R\rangle ,
\end{equation}
%%%%%%%%%%%%%%%%%%%%%%%%%%%%%%%%%%%%%%%%%%%%%%%%%%%%%%%%%%%%%%
\begin{equation}
\label{eq:ff}
[-\frac{1}{2\mu}\Delta + U_{fin;J'}(R)]
|fin,J',E';R\rangle = \epsilon'_{J',v'}\cdot|fin,J',E';R\rangle ,
\end{equation}
%%%%%%%%%%%%%%%%%%%%%%%%%%%%%%%%%%%%%%%%%%%%%%%%%%%%%%%%%%%%%%
where $\mu$ is the reduced mass of the considered ion-atom system,
%%%%%%%%%%%%%%%%%%%%%%%%%%%%%%%%%%%%%%%%%%%%%%%%%%%%%%%%%%%%%%
\begin{equation}
\label{eq:imp}
E'_{imp}=E'-(I_{A}-I_{B}),
\end{equation}
%%%%%%%%%%%%%%%%%%%%%%%%%%%%%%%%%%%%%%%%%%%%%%%%%%%%%%%%%%%%%%
and $\epsilon'_{J',v'} < 0$. With $U_{in;J}(R)$ and $U_{fin;J'}(R)$ are defined the effective
potential energies given by
%%%%%%%%%%%%%%%%%%%%%%%%%%%%%%%%%%%%%%%%%%%%%%%%%%%%%%%%%%%%%%
\begin{equation}
\label{eq:Ui}
\begin{array}{ll}
             \displaystyle{U_{in;J}(R)=U_{in}(R)+\frac{{\hbar}^{2} J(J+1)}{2 \mu R^{2}}}, \\
             \displaystyle{ U_{fin;J'}(R)=U_{fin}(R)+\frac{{\hbar}^{2} J'(J'+1)}{2 \mu R^{2}}},
             \end{array}
\end{equation}
%%%%%%%%%%%%%%%%%%%%%%%%%%%%%%%%%%%%%%%%%%%%%%%%%%%%%%%%%%%%%%
where $U_{in}(R)$ and $U_{fin}(R)$ are the adiabatic potential energies of the molecular
ions $AB^{+}$ and $(AB^{+})^{*}$ in the states $|in;R>$ and $|fin;R>$ as the functions of
$R$. In further consideration it is assumed that the radial wave functions which correspond
to the considered states satisfy the standard ortho-normalization conditions.

In the case $A=$ He and $B=$ H, as well as in the cases $A=$ H and $B=$ Mg or Al, each of the
groups (i) and (ii) of the electronic molecular states contains only one $\Sigma$-state:
the ground and the first excited electronic state of the considered molecular ion. Because
of that in these cases we will denote the states $|in; R>$ and $|fin; R>$ with $|1; R>$
and $|2; R>$ respectively, and the corresponding potential curves - with $U_{1}(R)$ and
$U_{2}(R)$. Let $D_{in;fin}(R)$ be the electronic dipole matrix element which corresponds
to the transitions given in Eqs. (\ref{eq:sheme}), i.e.
%%%%%%%%%%%%%%%%%%%%%%%%%%%%%%%%%%%%%%%%%%%%%%%%%%
\begin{equation}
\label{eq:De}
D_{in;fin}(R)=<in;R|{\bf D}(R)|fin;R>,
\end{equation}
%%%%%%%%%%%%%%%%%%%%%%%%%%%%%%%%%%%%%%%%%%%%%%%%%%
where $\bf{D}$ is the operator of the dipole moment of the considered ion-atom system. One
can see that in the mentioned cases
%%%%%%%%%%%%%%%%%%%%%%%%%%%%%%%%%%%%%%%%%%%%%%%%%%
\begin{equation}
\label{eq:D12}
D_{in;fin}(R) = D_{1;2}(R) \equiv <1;R|{\bf D}(R)|2;R>.
\end{equation}
%%%%%%%%%%%%%%%%%%%%%%%%%%%%%%%%%%%%%%%%%%%%%%%%%%%

\noindent However, in the case $A=$ H and $B=$ Si, the group (i) contains the ground
electronic $\Sigma$-state and the excited, weekly bounded $\Pi$-state, denoted here with
$|1a; R>$ and $|1b; R>$ respectively, while the group (ii) contains two $\Sigma$-states
and one $\Pi$-state, denoted here with $|2a; R>$, $|2b; R>$ and $|2c; R>$ respectively. In
accordance with this, the corresponding potential curves will be denoted by $U_{1a,1b}(R)$
and $U_{2a,2b,2c}(R)$. One can see that, in this case we have the situations when $|in,R> =
|1a; R>$ and  $|fin,R> = |2a; R>$ or $|2b; R>$, and $|in,R> = |1b; R>$ and $|fin,R> = |2c;
R>$. Consequently, the corresponding $D_{in;fin}(R)$ will be defined here by the relations
%%%%%%%%%%%%%%%%%%%%%%%%%%%%%%%%%%%%%%%%%%%%%%%%%%
\begin{equation}
\label{eq:Dabc}
D_{in;fin}(R) = \left\{
             \begin{array}{lll}
              D_{1a;2a}(R)\equiv<1a;R|{\bf D}(R)|2a;R>\\
               D_{1a;2b}(R)\equiv<1a;R|{\bf D}(R)|2b;R>\\
              D_{1b;2c}(R)\equiv<1b;R|{\bf D}(R)|2c;R>
             \end{array}
     \right.
\end{equation}
%%%%%%%%%%%%%%%%%%%%%%%%%%%%%%%%%%%%%%%%%%%%%%%%%%%
For the ions HeH$^{+}$ and (HeH$^{+})^{*}$ the potential curves $U_{1,2}(R)$ and the values of
$D_{1;2}(R)$ are taken from \cite{gre74a,gre74b}. For all other considered molecular ions the
corresponding potential curves and the values of the dipole matrix elements are calculated
within this work. Also, we will introduce the so called splitting term $U_{in;fin}(R)$,
defined by
%%%%%%%%%%%%%%%%%%%%%%%%%%%%%%%%%%%%%%%%%%%%%%%%%%%
\begin{equation}
\label{eq:U12}
U_{in;fin}(R) = U_{fin}(R)-U_{in}(R),
\end{equation}
%%%%%%%%%%%%%%%%%%%%%%%%%%%%%%%%%%%%%%%%%%%%%%%%%%%
which is used in the further considerations.

\begin{figure}
\begin{center}
\includegraphics[width=\columnwidth,
height=0.75\columnwidth]{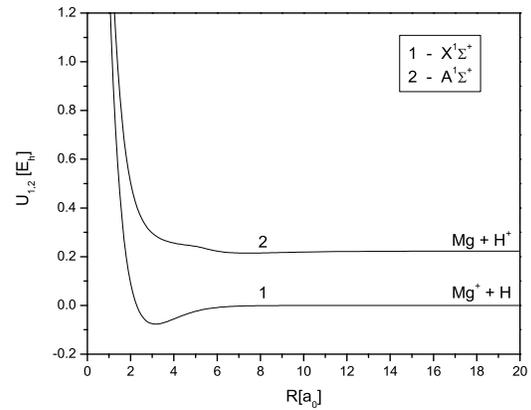} \caption{The potential curves of the molecular ion
HMg$^{+}$.}
\label{fig:HMg_term}
\end{center}
\end{figure}

\begin{figure}
\begin{center}
\includegraphics[width=\columnwidth,
height=0.75\columnwidth]{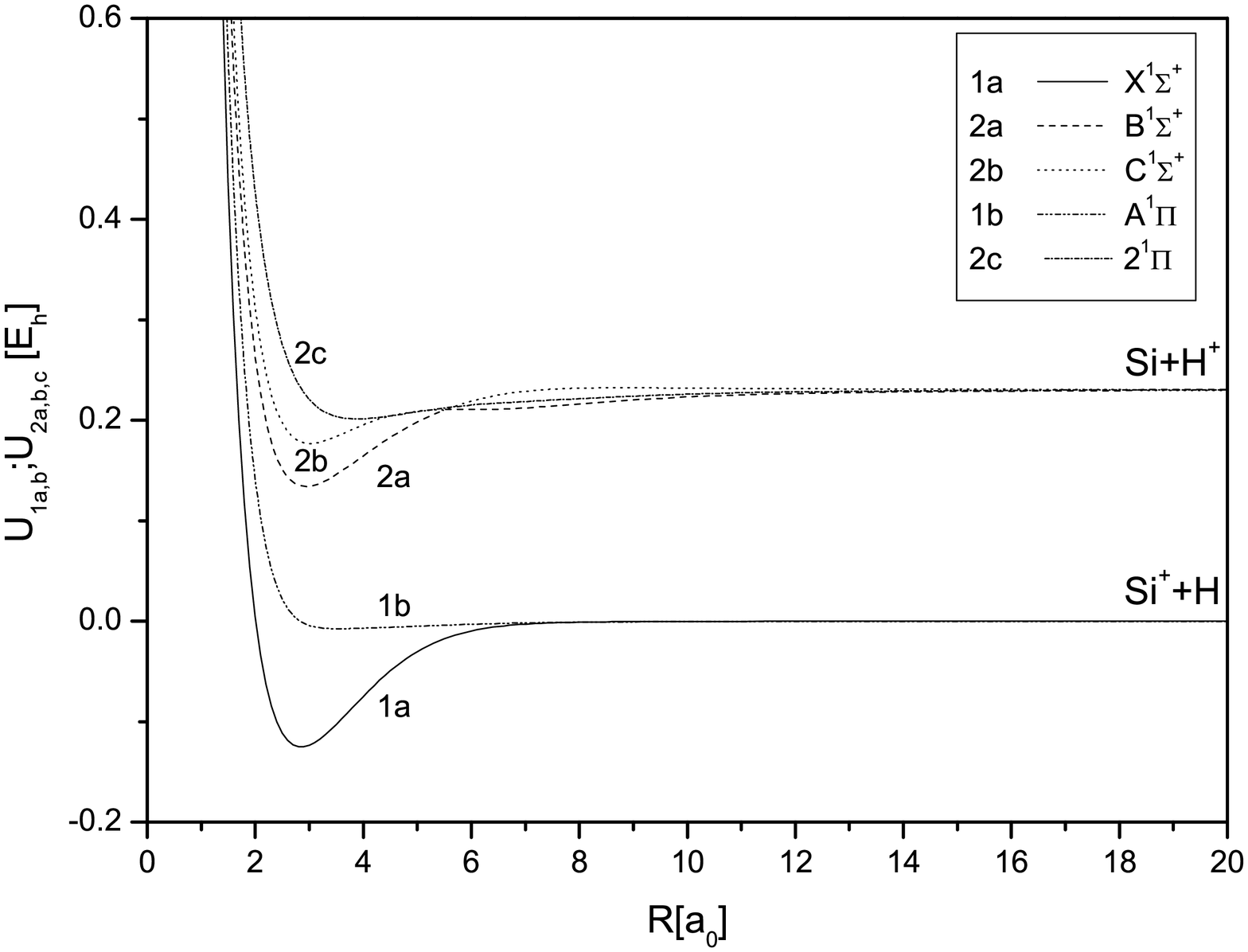} \caption{Same as in Fig.\ref{fig:HMg_term}, but for
the molecular ion HSi$^{+}$.}
\label{fig:HSi_term}
\end{center}
\end{figure}

\begin{figure}
\begin{center}
\includegraphics[width=\columnwidth,
height=0.75\columnwidth]{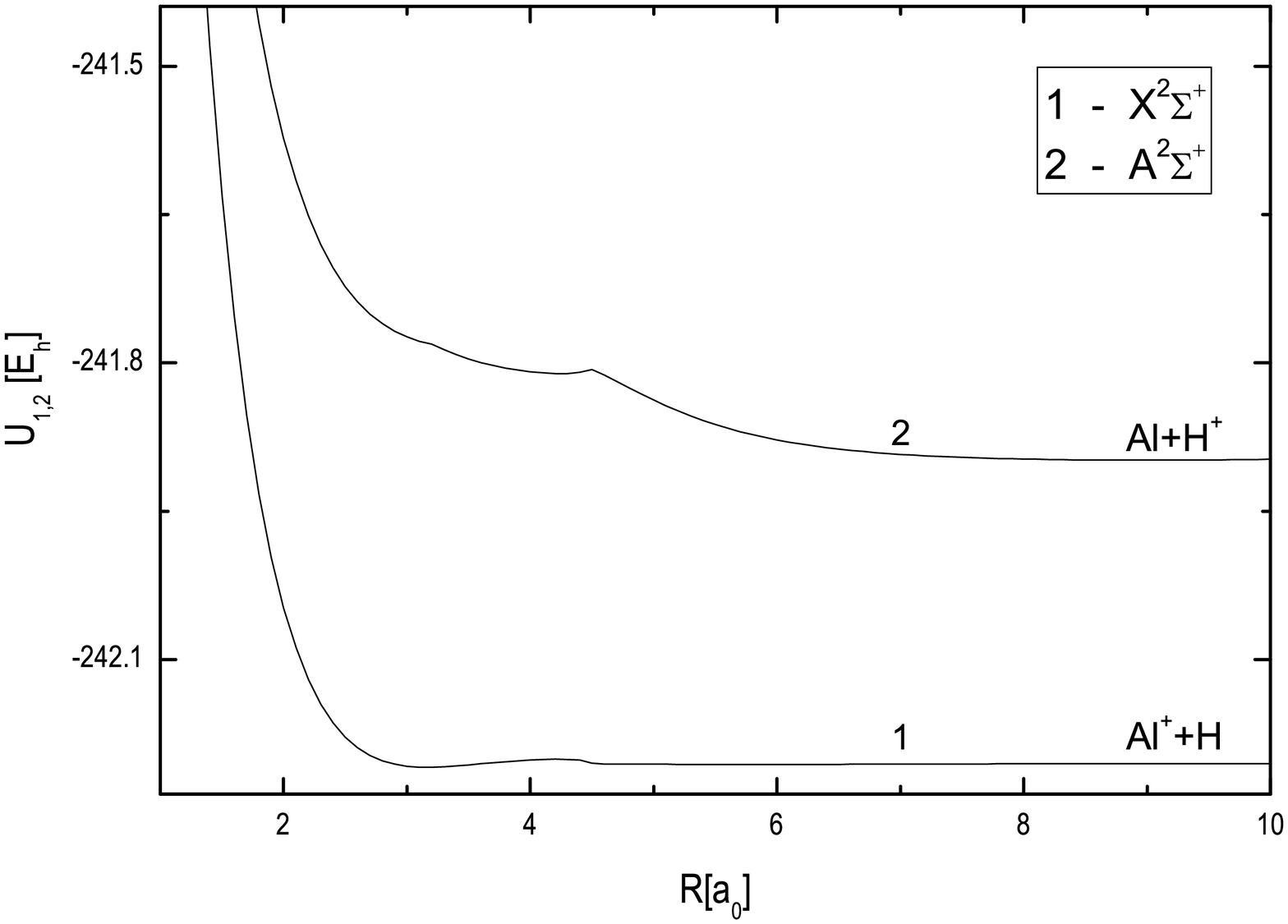} \caption{Same as in Fig.\ref{fig:HMg_term}, but
for the molecular ion
HAl$^{+}$ .}
\label{fig:HAl_term}
\end{center}
\end{figure}

\begin{figure}
\begin{center}
\includegraphics[width=\columnwidth,
height=0.75\columnwidth]{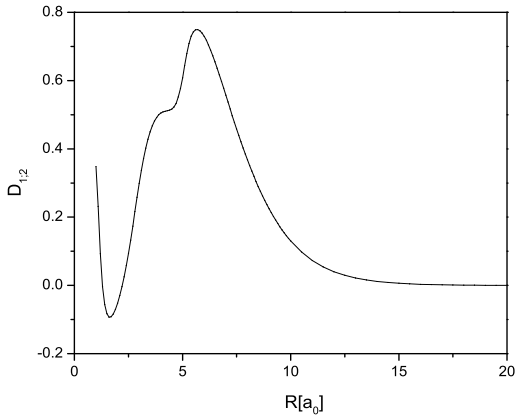} \caption{The behaviour of the electronic dipole matrix
element $D_{1;2}(R)$, given by equations (\ref{eq:De}) and (\ref{eq:D12}), for the molecular ion
HMg$^{+}$.}
\label{fig:HMg_dip}
\end{center}
\end{figure}

\begin{figure}
\begin{center}
\includegraphics[width=\columnwidth,
height=0.75\columnwidth]{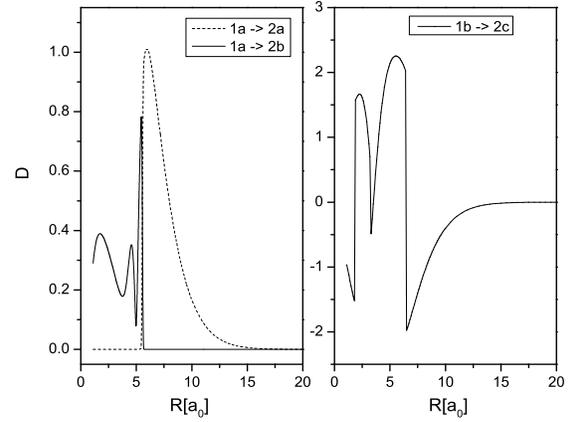} \caption{The behavior of dipole matrix elements
$D_{1a;2a;b}(R)$ and $D_{1b;2c}(R)$, given by equations (\ref{eq:De}) and (\ref{eq:Dabc}),
for the molecular ion HSi$^{+}$.}
\label{fig:HSi_dip}
\end{center}
\end{figure}

\begin{figure}
\begin{center}
\includegraphics[width=\columnwidth,
height=0.75\columnwidth]{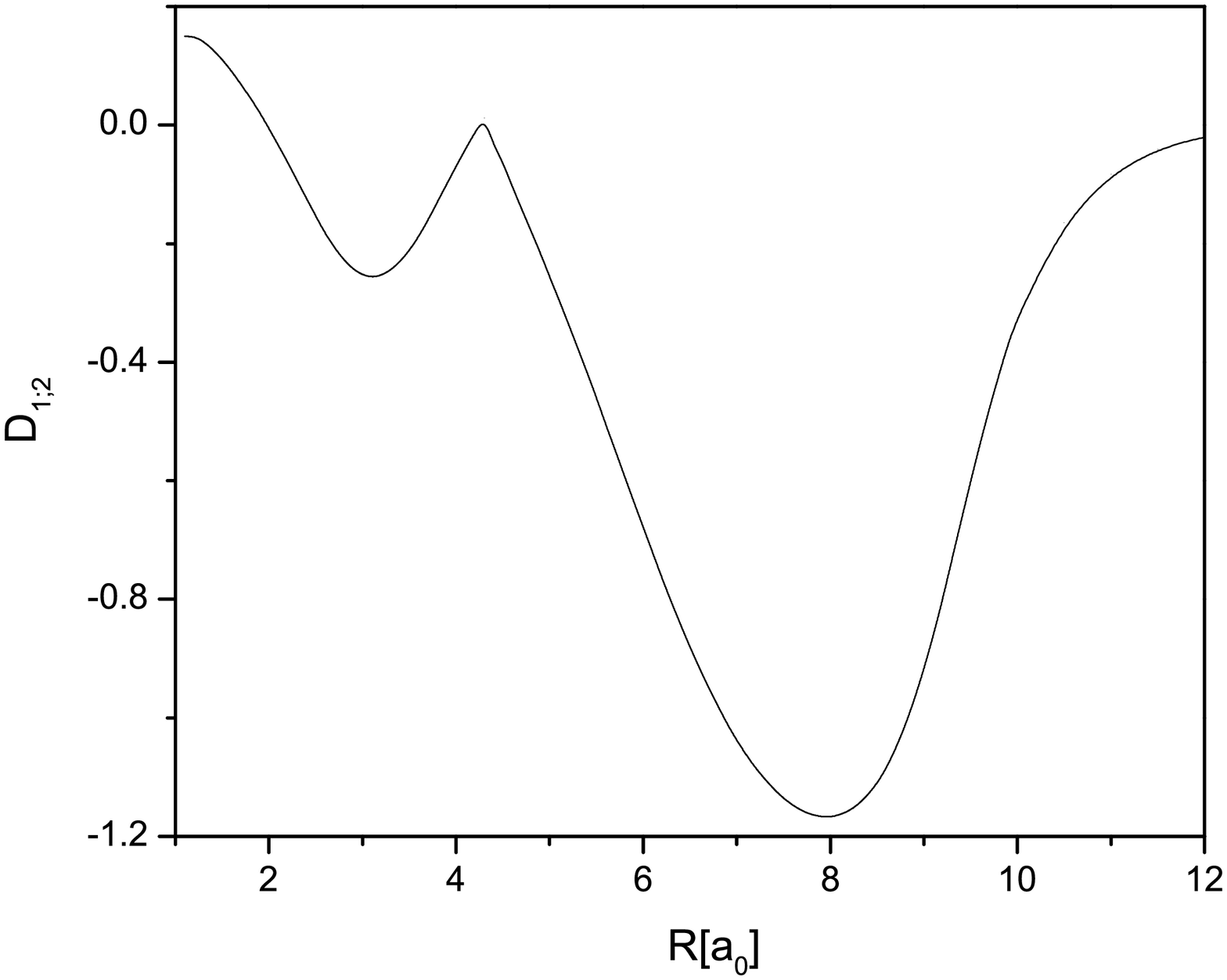} \caption{Same as in Fig.\ref{fig:HMg_dip}, but for the
molecular ion HAl$^{+}$.}
\label{fig:HAl_dip}
\end{center}
\end{figure}

All calculations were done at the multi-configuration self-consistent field (MCSCF) and
multi-reference configuration interaction (MRCI) levels (with MCSCF orbitals) using the MOLPRO
package of programs \citep{mol06}. The basis sets we employed were the cc-pvqz basis sets of
Dunning et al. \citep{dun89,ken92}. Initially, test runs were done in the asymptotic region
(30 Bohr) at the lowest three to five  levels at each selected symmetry to determine the low
lying levels corresponding to the H-$B^{+}$ and the H$^{+}$-$B$ electron distributions and
their wave functions ($B =$ Mg, Si, Al). For each level a Mulliken population analysis was
done to determine the location of the charges. Since H has the highest ionization potential,
the lowest states of a given symmetry correspond to the H-$B^{+}$ charge distribution while
the H$^{+}$-$B$ charge distribution is described by one or more of the excited states. The
potential energies of all these states were calculated starting at the asymptotic region
and moving inwards. At each point the dipole matrix elements between the H-$B^{+}$ and
H$^{+}$-$B$ states were calculated using the corresponding wave functions. The calculated
potential curves are presented in Figs. \ref{fig:HMg_term}-\ref{fig:HAl_term}.  Figures
\ref{fig:HMg_dip}-\ref{fig:HAl_dip} show the behavior of $D_{1;2}(R)$, $D_{1a;2a;b}(R)$
and $D_{1b;2c}(R)$.

%%%%%%%% termovi %%%%%%
%%%%%%%%%%%%%%%%%%%%%%%%
%%%%%%%%%%%%%%%%%%%%%%%%%
%%%%%%%%%%%%%%%%%%%%%%%%%%%%%%%%%%%%%%%%%%%%%%%%%%%
%%%%%% dipoli %%%%%%%%
%%%%%%%%%%%%%%%%%%%%%%%
%%%%%%%%%%%%%%%%%%%%%%%%%
%%%%%%%%%%%%%%%%%%%%%%%%%%%%%%%%%%%%%%%%%%%%%%%%%%%%%
\section {The relevant spectral characteristics}

In accordance with the aim of this work the considered absorption processes will be
characterized by the adequately defined spectral absorption coefficients. We will start
from the bound-free, free-free and free-bound absorption processes, caused by the radiative
transitions given by Eqs.(\ref{eq:sheme}), for the given species $A$ and $B$. In
the cases $A=$ H and $B=$ Mg and $A=$ He and $B=$ H where the transitions given by
Eqs.(\ref{eq:sheme}) are connected with only one initial and one final
$\Sigma$-electronic state, the corresponding spectral absorption coefficients are
denoted here with $\kappa^{(bf)}_{AB^{+}}(\lambda, T)$,
$\kappa^{(ff)}_{AB^{+}}(\lambda, T)$ and $\kappa^{(fb)}_{AB^{+}}(\lambda, T)$,
where $T$ is the local plasma temperature in the solar atmosphere.

However, it follows from the above mentioned that in the case $A=$ H and $B=$ Si we will
have the transitions from two initial $\Sigma$- and $\Pi$-electronic states to two final
$\Sigma$- and one final $\Pi$-electronic states. Because of that in this case we will have
three groups of the spectral absorption coefficients $\kappa^{(bf,ff,fb)}_{AB^{+}}(\lambda,
T;i,f)$, which correspond to these transitions.

\subsection {The bound-free processes.} In the usual way the spectral absorption coefficients
$\kappa_{bf}(\lambda, T)$, which characterize the efficiency of the photo-dissociation
process (\ref{eq:nonsim1}) are defined by
%%%%%%%%%%%%%%%%%%%%%%%%%%%%%%%%%%%%%%%%%%%%%%%%%%%%%%%%%%%%%%%%%%
\begin{equation}
\label{eq:phd0}
\kappa^{(bf)}_{AB^{+}}(\lambda, T)= \sigma^{(phd)}_{AB^{+}}(\lambda,T)\cdot N_{AB^{+}},
\end{equation}
%%%%%%%%%%%%%%%%%%%%%%%%%%%%%%%%%%%%%%%%%%%%%%%%%%%%%%%%%%%%%%%%%%
where $N(AB^{+})$ is the local density of the considered molecular ion $AB^{+}$, and
$\sigma_{phd}$ is the corresponding mean thermal photo-dissociation cross section, which
is given by
%%%%%%%%%%%%%%%%%%%%%%%%%%%%%%%%%%%%%%%%%%%%%%%%%%%%%%%%%%%%%%%%%%
\begin{equation}
\label{eq:sigbf}
\sigma^{(phd)}_{AB^{+}}(\lambda,T)=\frac{\sum\limits_{J,v} (2J+1)
e^{\frac{-E_{J,v}}{kT}}\cdot \sigma_{J,v}(\lambda) }
{\sum\limits_{J,v}(2J+1) e^{\frac{-E_{J,v}}{kT}}},
\end{equation}
%%%%%%%%%%%%%%%%%%%%%%%%%%%%%%%%%%%%%%%%%%%%%%%%%%%%%%%%%%%%%%%%%%
where $\sigma_{J,v}(\lambda)$ is the partial photo-dissociation cross-section for the
ro-vibrational states with given quantum numbers $J$ and $v$, and $E_{J,v}$ - the energies
of these states with the respect to the ground ro-vibrational states. It means that $E_{J,v}
= E_{dis} + \epsilon_{J,v}$, where $E_{dis}$ is the dissociative energy of the ion $AB^{+}$,
and the energies $\epsilon_{J,v}<0$ are determined from  Eq.(\ref{eq:b}) together with the
wave functions of the considered ro-vibrational states. Within the dipole approximation
the partial cross-sections $\sigma_{J,v}(\lambda)$ are given by the expressions
%%%%%%%%%%%%%%%%%%%%%%%%%%%%%%%%%%%%%%%%%%%%%%%%%%%%%%%%%%%%%%%%%%
\begin{equation}
\label{eq:sigvJ}
\begin{split}
\sigma_{J,v}(\lambda)=\frac{8 \pi^{3}}{3 \lambda }
[\frac{J+1}{2J+1}|D_{J,v;J+1,E'_{imp}}|^{2}\\
+ \frac{J}{2J+1}|D_{J,v;J-1,E'_{imp}}|^{2}],
\end{split}
\end{equation}
%%%%%%%%%%%%%%%%%%%%%%%%%%%%%%%%%%%%%%%%%%%%%%%%%%%%%%%%%%%%%%%%%%
%%%%%%%%%%%%%%%%%%%%%%%%%%%%%%%%%%%%%%%%%%%%%%%%%%%%%%%%%%%%%%%%%%
\begin{equation}
D_{J,v;J \pm 1,E'_{imp}} =
<in,J,v; R|D_{in,fin}(R)|fin,J\pm 1,E'>,
\label{eq:D}
\end{equation}
%%%%%%%%%%%%%%%%%%%%%%%%%%%%%%%%%%%%%%%%%%%%%%%%%%%%%%%%%%%%%%%%%%
where $E' = \epsilon_{J,v} + \varepsilon_{\lambda}$, $E'_{imp}$ and $E'$ are connected with
Eq.(\ref{eq:imp}), and $D_{in,fin}(R)$ is given by Eqs. (\ref{eq:De}) - (\ref{eq:Dabc}).

Keeping in mind that the deviations from the local thermodynamical equilibrium (LTE) of
the used model C from \citet{ver81} are not related to the considered bound-free processes,
we will take the photo-dissociation coefficient $\kappa^{(bf)}_{AB^{+}}(\lambda, T)$ in an
equivalent form, suitable for further considerations, namely
%%%%%%%%%%%%%%%%%%%%%%%%%%%%%%%%%%%%%%%%%%%%%%%%%%%%%%%%%%%%%%%%%%
\begin{equation}
\kappa^{(bf)}_{AB^{+}}(\lambda, T)=K^{(bf)}_{AB^{+}}(\lambda,T) \cdot N_{A}N_{B^{+}},
\label{eq:kapaiabf}
\end{equation}
%%%%%%%%%%%%%%%%%%%%%%%%%%%%%%%%%%%%%%%%%%%%%%%%%%%%%%%%%%%%%%%%%%
%%%%%%%%%%%%%%%%%%%%%%%%%%%%%%%%%%%%%%%%%%%%%%%%%%%%%%%%%%%%%%%%%%
\begin{equation}
K^{(bf)}_{AB^{+}}(\lambda,T)=\sigma^{(phd)}_{AB^{+}}(\lambda,T) \cdot \chi^{-1}(T;AB^{+}),
\end{equation}
\begin{equation}
\label{eq:Kiaa}
\chi(T;AB^{+})=\left [\frac{N(A)N(B^{+})}{N(AB^{+})} \right],
\end{equation}
%%%%%%%%%%%%%%%%%%%%%%%%%%%%%%%%%%%%%%%%%%%%%%%%%%%%%%%%%%%%%%%%%%
where the factor $\chi$ is given by the relation
%%%%%%%%%%%%%%%%%%%%%%%%%%%%%%%%%%%%%%%%%%%%%%%%%%%%%%%%%%%%%%%%%%
\begin{equation}
\label{eq:mass}
\chi(T;AB^{+})=\frac{g_{A}g_{B^{+}}}{g_{AB^{+}}}\left(\frac{\mu kT}{2\pi
\hbar^{2}}\right)^{\frac{3}{2}}\cdot
\frac{1}{\sum\limits_{J,v} (2J+1)
e^{\frac{E_{dis}-E_{J,v}}{kT}}},
\end{equation}
%%%%%%%%%%%%%%%%%%%%%%%%%%%%%%%%%%%%%%%%%%%%%%%%%%%%%%%%%%%%%%%%%%
where $g_{AB^{+}}$, $g_{A}$ and $g_{B^{+}}$ are the electronic statistical weights of the
species $AB^{+}$, $A$ and $B^{+}$ respectively, and $\sigma^{(phd)}_{AB^{+}}(\lambda,T)$
is given by Eqs.~(\ref{eq:sigbf})-(\ref{eq:D}). The behavior of the photo-dissociation
cross section $\sigma^{(phd)}_{AB^{+}}(\lambda,T)$ and the bound-free spectral rate
coefficient $K^{(bf)}_{AB^{+}}(\lambda,T)$ are illustrated in Figs. \ref{fig:HMg_sigphd}
and \ref{fig:HMg_Kbf}, on the example of the case $A=$ H and $B=$ Mg, for 110 nm $\lesssim
\lambda \lesssim 205$ nm and $T=4000$ K, $T=6000$ K, $T=8000$ K and $T=10000$ K. These
figures show that exist a significant difference between temperature dependence of the
mean thermal photo-ionization cross section and the corresponding spectral rate coefficient.

\subsection {The free-free processes.} The very fast approaching of the electronic dipole matrix
elements $D_{in,fin}(R)$ to zero with the increasing of $R$ in the case of the non-symmetric
ion-atom systems, which is illustrated by Figs. \ref{fig:HMg_dip}-\ref{fig:HAl_dip},
makes possible to apply here the complete quantum mechanical treatment not
only to the bound-free and free-bound absorption processes (\ref{eq:nonsim1}) and
(\ref{eq:nonsim3}), but to the free-free absorption process (\ref{eq:nonsim2}). Namely,
it can be shown (see for an example \cite{leb02}) that the free-free spectral absorption
coefficients $\kappa^{(ff)}_{AB^{+}}(\lambda, T)$ can be expressed over the quantities
$\sigma^{(ff)}_{AB^{+}}(\lambda,E) \equiv \sigma^{(ff)}_{AB^{+}}(J,E,\lambda;J \pm 1,E'_{imp})$
in the form
%%%%%%%%%%%%%%%%%%%%%%%%%%%%%%%%%%%%%%%%%%%%%%%%%%%%%%%%%%%%%%%%%%
\begin{equation}
\label{eq:kapaff}
\begin{array}{ll}
             \displaystyle{\kappa^{(ff)}_{AB^{+}}(\lambda,T) = K^{(ff)}_{AB^{+}}(\lambda,T) \cdot N_{A}N_{B^{+}},}\\
             \displaystyle{ K^{(ff)}_{AB^{+}}(\lambda,T)=\int\limits_{0}^{\infty} \left(\frac{2E}{\mu} \right)^{\frac{1}{2}} \sigma^{(ff)}_{AB^{+}}(\lambda,E) f_{T}(E) dE},
             \end{array}
\end{equation}
%%%%%%%%%%%%%%%%%%%%%%%%%%%%%%%%%%%%%%%%%%%%%%%%%%%%%%%%%%%%%%%%%%
\noindent where $f_{T}(E)$ is the Maxwell impact energy distribution function
%%%%%%%%%%%%%%%%%%%%%%%%%%%%%%%%%%%%%%%%%%%%%%%%%%%%%%%%%%%%%%%%%%
\begin{equation}
f_{T}(E)=\frac{2}{\pi^{1/2} (kT)^{3/2}} e^{-\frac{E}{k_{B}T}} E^{1/2} dE,
\label{eq:fT}
\end{equation}
%%%%%%%%%%%%%%%%%%%%%%%%%%%%%%%%%%%%%%%%%%%%%%%%%%%%%%%%%%%%%%%%%%
and $\sigma^{(ff)}_{AB^{+}}(\lambda,E)$ is given by
%%%%%%%%%%%%%%%%%%%%%%%%%%%%%%%%%%%%%%%%%%%%%%%%%%%%%%%%%%%%%%%%%%
\begin{equation}
\label{eq:sigff}
\begin{split}
\sigma^{(ff)}_{AB^{+}}(\lambda)&=\frac{g_{A^{+}}g_{B}}{g_{A}g_{B^{+}}}\frac{8 \pi^{4} \hbar
^{2} \varepsilon_{\lambda}}{3 c \cdot 2 \mu E }\cdot\\
&\left[(J+1)\cdot|D_{J,E;J+1,E'_{imp}}|^{2}+
J\cdot|D_{J,E;J-1,E'_{imp}}|^{2} \right],
\end{split}
\end{equation}
%%%%%%%%%%%%%%%%%%%%%%%%%%%%%%%%%%%%%%%%%%%%%%%%%%%%%%%%%%%%%%%%%%
%%%%%%%%%%%%%%%%%%%%%%%%%%%%%%%%%%%%%%%%%%%%%%%%%%%%%%%%%%%%%%%%%%
\begin{equation}
D_{J,E;J \pm 1,E'_{imp}} =
<in,J,E; R|D_{in,fin}(R)|fin,J\pm 1,E'>,
\label{eq:D}
\end{equation}
%%%%%%%%%%%%%%%%%%%%%%%%%%%%%%%%%%%%%%%%%%%%%%%%%%%%%%%%%%%%%%%%%%
where $E'_{imp}$ and $E'= E + \varepsilon_{\lambda}$ are connected with Eq.(\ref{eq:imp}),
$g_{A^{+}}$, $g_{B}$, $g_{A}$ and $g_{B^{+}}$ are the electronic statistical weights of
the species $A^{+}$, $B$, $A$ and $B^{+}$ respectively. One can see that the quantity
$\sigma^{(ff)}_{AB^{+}}$ can be treated as the effective cross section, but is expressed
in units cm$^{4}$s, and the rate coefficient $K^{(ff)}_{AB^{+}}(\lambda,T)$ is equal to
the absorption coefficient for the unit densities  $N(A)$ and $N(B^{+})$.

\subsection {The free-bound processes.} Similarly to the free-free case the free-bound spectral
absorption coefficients $\kappa^{(fb)}_{AB^{+}}(\lambda, T)$ is taken here as
%%%%%%%%%%%%%%%%%%%%%%%%%%%%%%%%%%%%%%%%%%%%%%%%%%%%%%%%%%%%%%%%%%
\begin{equation}
\label{eq:Kfb0}
\kappa^{(fb)}_{AB^{+}}(\lambda, T) = K^{(fb)}_{AB^{+}}(\lambda,T) \cdot N_{A}N_{B^{+}},
\end{equation}
%%%%%%%%%%%%%%%%%%%%%%%%%%%%%%%%%%%%%%%%%%%%%%%%%%%%%%%%%%%%%%%%%%
where the rate coefficient $K^{(fb)}_{AB^{+}}(\lambda,T)$ can be also expressed over the
corresponding free-bound cross section. In accordance with \citet{mih96} and \citet{ign99}
it can be presented in the form
%%%%%%%%%%%%%%%%%%%%%%%%%%%%%%%%%%%%%%%%%%%%%%%%%%%%%%%%%%%%%%%%%%
\begin{equation}
\label{eq:Kfb}
K^{(fb)}_{AB^{+}}(\lambda,T)= \frac{(2 \pi)^{3}}{3 \hbar \lambda} \left(\frac{2
\pi\hbar^{2}}{\mu kT}\right)^{3/2}
\sum_{J',v'}\left(\frac{\mu}{2E}\right)^{1/2}e^{-\frac{E}{kT}}\cdot C_{J',v'},
\end{equation}
%%%%%%%%%%%%%%%%%%%%%%%%%%%%%%%%%%%%%%%%%%%%%%%%%%%%%%%%%%%%%%%%%%
\begin{equation}
\label{eq:Cfb}
C_{J',v'}=\frac{g_{(AB^{+})^{*}}}{g_{A}g_{B^{+}}} \cdot
[J'|D_{J'-1,E;J',v'}|^{2}+(J'+1)|D_{J'+1,E;J',v'}|^{2}],
\end{equation}
%%%%%%%%%%%%%%%%%%%%%%%%%%%%%%%%%%%%%%%%%%%%%%%%%%%%%%%%%%%%%%%%%%
\begin{equation}
\label{eq:Dfb}
D_{J' \pm 1,E;J',v'} =
<in,J' \pm 1,E; R |D_{in,fin}(R)|fin, J',v'>,
\end{equation}
%%%%%%%%%%%%%%%%%%%%%%%%%%%%%%%%%%%%%%%%%%%%%%%%%%%%%%%%%%%%%%%%%%
\noindent where $E = I_{A}-I_{B}-\varepsilon_{\lambda} + \epsilon_{J';v'}$, $g_{(AB^{+})^{*}}$
is the electronic statistical weights of the molecular ion $(AB^{+})^{*}$, $\epsilon_{J';v'}<0$
-the energy of the ion $(AB^{+})^{*}$ in the ro-vibrational state with the orbital
and vibrational quantum numbers $J'$ and $v'$, and summing is performed over all these
ro-vibration states. Let us note that within this paper we will neglect everywhere the corrections
for stimulated emission as in all the cases considered here the corresponding corrections
(given the relevant values of the ratio $\varepsilon_{\lambda}/kT$) would be at a level of $0.01\%$.

The behavior of the free-free and free-bound spectral rate coefficients
$K^{(ff)}_{AB^{+}}(\lambda,T)$ and $K^{(fb)}_{AB^{+}}(\lambda,T)$ is illustrated by the
Figs. \ref{fig:HeH_Kfffb}-\ref{fig:HSi_Kfffb}, on the examples:\\
\noindent $A=$ He and $B=$ H for 60 nm$ \lesssim \lambda \lesssim $115 nm and $T=4000$ K,
6000 K and 8000 K;\\
$A=$ H and $B=$ Mg for 150 nm$ \lesssim \lambda \lesssim $220nm and $T=4000$ K, 6000 K,
8000 K and 10000 K; \\
$A=$ H and $B=$ Si, for the transition $X^{1}\Sigma^{+}\rightarrow B^{1}\Sigma^{+}$, for 190
nm $\lesssim \lambda \lesssim$ 220 nm and $T=4000$ K, 6000 K, 8000 K and 10000 K.\\

These figures show that in the general case the absorption processes caused by the free-free
and free-bound transitions (\ref{eq:sheme}) have to be considered together since their relative
efficiency, characterized by $K^{(ff)}_{AB^{+}}(\lambda,T)$ and $K^{(fb)}_{AB^{+}}(\lambda,T)$
significantly changes from one to other ion-atom system. Let us note that at least some of
the picks, existing in Figs. \ref{fig:HeH_Kfffb}-\ref{fig:HSi_Kfffb} which illustrate the
shape of the profiles $K^{(ff)}_{AB^{+}}(\lambda,T)$ and $K^{(fb)}_{AB^{+}}(\lambda,T)$,
can be connected with the extremums of the corresponding splitting terms. Such phenomena
in connection with the ion-atom systems were discussed already in \cite{mih81a}. Let us
note that in the case of non-symmetric atom-atom systems the similar phenomena were also
investigated earlier \citep{vez98,ske02}.

\subsection {The partial and total non-symmetric spectral absorption coefficients.} The partial
absorption coefficients which characterize the individual contribution of the considered
ion-atom systems are denoted here
with $\kappa_{AB^{+}}(\lambda) \equiv \kappa_{AB^{+}}(\lambda,T; N_{A},N_{B^{+}})$. In
accordance with the above mentioned we have that
%%%%%%%%%%%%%%%%%%%%%%%%%%%%%%%%%%%%%%%%%%%%%%%%%%%%%%%%%%%%%%%%%%
\begin{equation}
\label{eq:kapa}
\kappa_{AB^{+}}(\lambda)=\kappa^{(bf)}_{AB^{+}}(\lambda,T) + \kappa^{(ff)}_{AB^{+}}(\lambda,T)
+ \kappa^{(fb)}_{AB^{+}}(\lambda,T),
\end{equation}
%%%%%%%%%%%%%%%%%%%%%%%%%%%%%%%%%%%%%%%%%%%%%%%%%%%%%%%%%%%%%%%%%%
\noindent in the cases $A=$ He and $B=$ H, and $A=$ H and $B=$ Mg and Al,
$\kappa_{AB^{+}}(\lambda)$, and that
%%%%%%%%%%%%%%%%%%%%%%%%%%%%%%%%%%%%%%%%%%%%%%%%%%%%%%%%%%%%%%%%%%
\begin{equation}
\label{eq:kapaSi}
\kappa_{\textrm{HSi}^{+}}(\lambda) = \frac{1}{3}[\kappa_{\textrm{HSi}^{+};1}(\lambda; \Sigma) +
\kappa_{\textrm{HSi}^{+};2}(\lambda; \Sigma)] + \frac{2}{3}\kappa_{\textrm{HSi}^{+}}(\lambda;
\Pi),
\end{equation}
%%%%%%%%%%%%%%%%%%%%%%%%%%%%%%%%%%%%%%%%%%%%%%%%%%%%%%%%%%%%%%%%%%
\noindent where $\kappa_{\textrm{HSi}^{+};1}(\lambda; \Sigma)$ and $\kappa_{\textrm{HSi}^{+};2}(\lambda;
\Sigma)$ describe the contribution of the radiative transitions $X^{1}\Sigma^{+} \rightarrow
B^{1}\Sigma^{+}$ and $X^{1}\Sigma^{+} \rightarrow C^{1}\Sigma^{+}$ respectively, and
$\kappa_{\textrm{HSi}^{+}}(\lambda; \Pi)$ - the contribution of the transition $A^{1}\Pi \rightarrow
2^{1}\Pi$. The spectral absorption coefficients $\kappa^{(bf)}_{AB^{+}}(\lambda,T)$,
$\kappa^{(ff)}_{AB^{+}}(\lambda,T)$ and $\kappa^{(fb)}_{AB^{+}}(\lambda,T)$ are defined by
Eqs. \ref{eq:phd0}-\ref{eq:mass}, \ref{eq:kapaff}-\ref{eq:D} and \ref{eq:Kfb0}-\ref{eq:Dfb}
respectively.

The total contribution of the mentioned non-symmetric ion-atom absorption processes to the
opacity of the considered stelar atmospheres within this work is described by the spectral
absorption coefficient $\kappa_{ia;nsim}(\lambda) \equiv \kappa_{ia;nsim}(\lambda;T)$, given by
%%%%%%%%%%%%%%%%%%%%%%%%%%%%%%%%%%%%%%%%%%%%%%%%%%%%%%%%%%%%%%%%%%
\begin{equation}
\label{eq:kapatot}
\kappa_{ia;nsim}(\lambda) = \sum \kappa_{AB^{+}}(\lambda),
\end{equation}
%%%%%%%%%%%%%%%%%%%%%%%%%%%%%%%%%%%%%%%%%%%%%%%%%%%%%%%%%%%%%%%%%%
\noindent where the partial coefficients $\kappa_{AB^{+}}(\lambda;T)$ are given by
Eqs. (\ref{eq:kapa}) and (\ref{eq:kapaSi}), and summing is performed over all considered
pairs of atom and ion species ($A$, $B^{+}$).  It is assumed that these coefficients are
determined with the plasma temperature $T$ and the atom and ion densities  taken from the
used model of the considered atmosphere.
%%%%%%% sigma & K %%%%%
\begin{figure}
\begin{center}
\includegraphics[width=\columnwidth,
height=0.75\columnwidth]{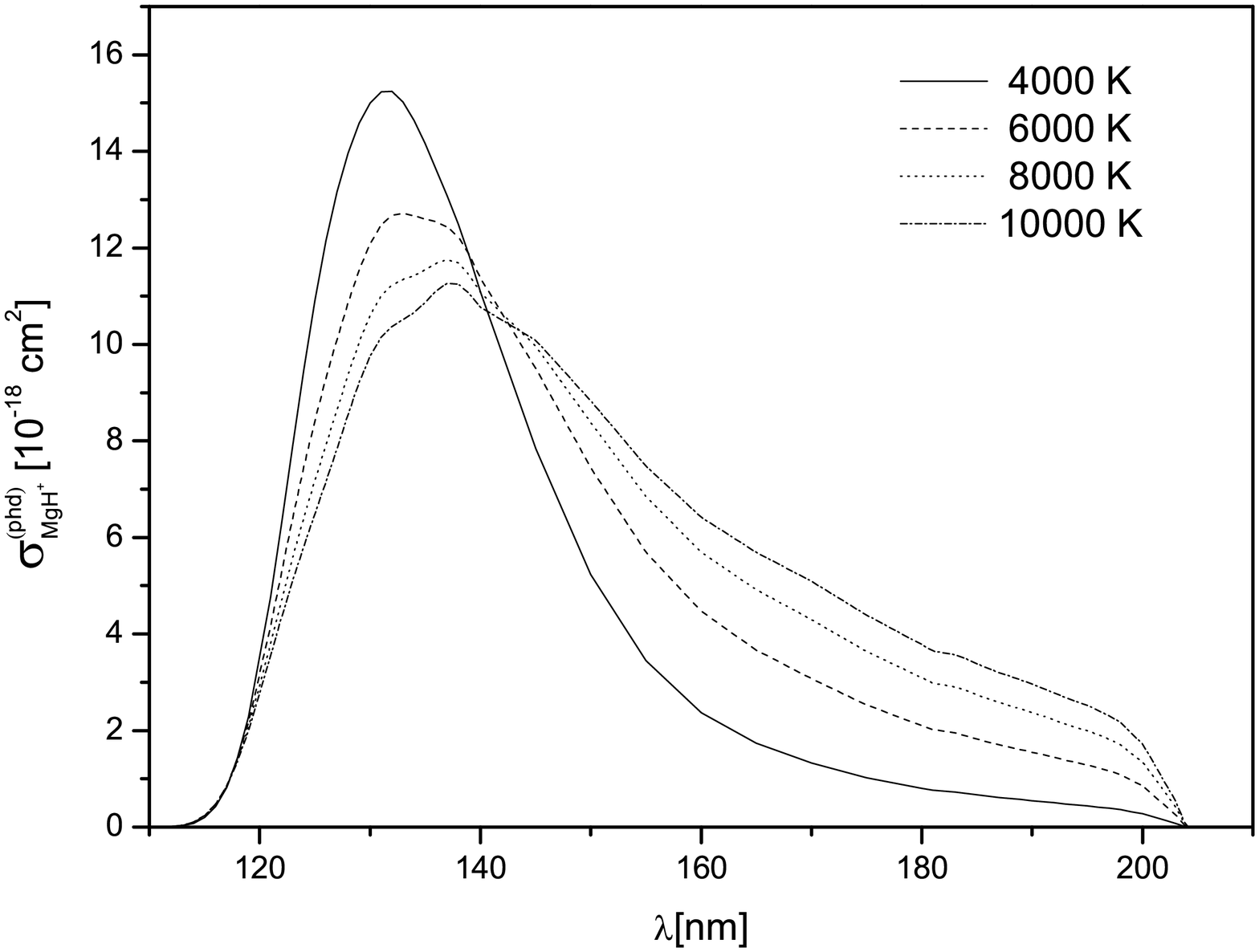} \caption{The behaviour of the mean thermal
photodissociation cross-section
$\sigma^{(phd)}_{HMg^{+}}(\lambda;T)$ for the molecular ion HMg$^{+}$.}
\label{fig:HMg_sigphd}
\end{center}
\end{figure}
%%%%%%%%%%%%%%%%%%%%%%%
\begin{figure}
\begin{center}
\includegraphics[width=\columnwidth,
height=0.75\columnwidth]{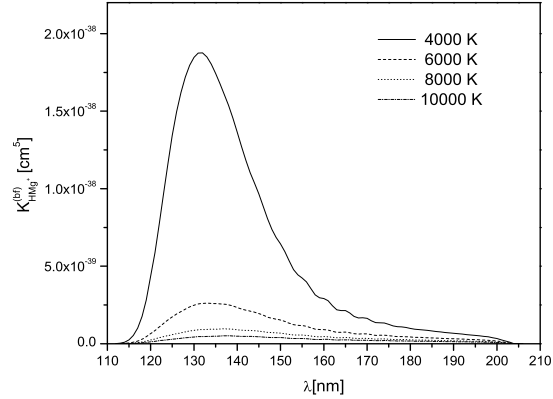} \caption{The behaviour of the bound-free (bf) spectral
rate coefficient $K^{(bf)}_{HMg^{+}}(\lambda;T)$ for the molecular ion HMg$^{+}$.}
\label{fig:HMg_Kbf}
\end{center}
\end{figure}
%%%%%%%%%%%%%%%%%%%%%%%
\begin{figure}
\begin{center}
\includegraphics[width=\columnwidth,
height=0.75\columnwidth]{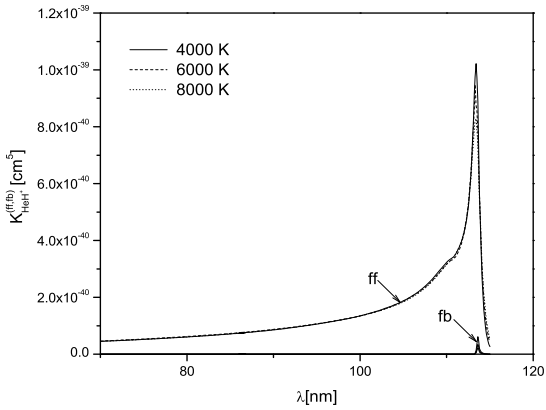} \caption{The behavior of the free-free (ff) and
free-bound (fb) spectral rate coefficients
$K^{(ff,fb)}(\lambda;T)$ for HeH$^{+}$.}
\label{fig:HeH_Kfffb}
\end{center}
\end{figure}
%%%%%%%%%%%%%%%%%%%%%%%%
\begin{figure}
\begin{center}
\includegraphics[width=\columnwidth,
height=0.75\columnwidth]{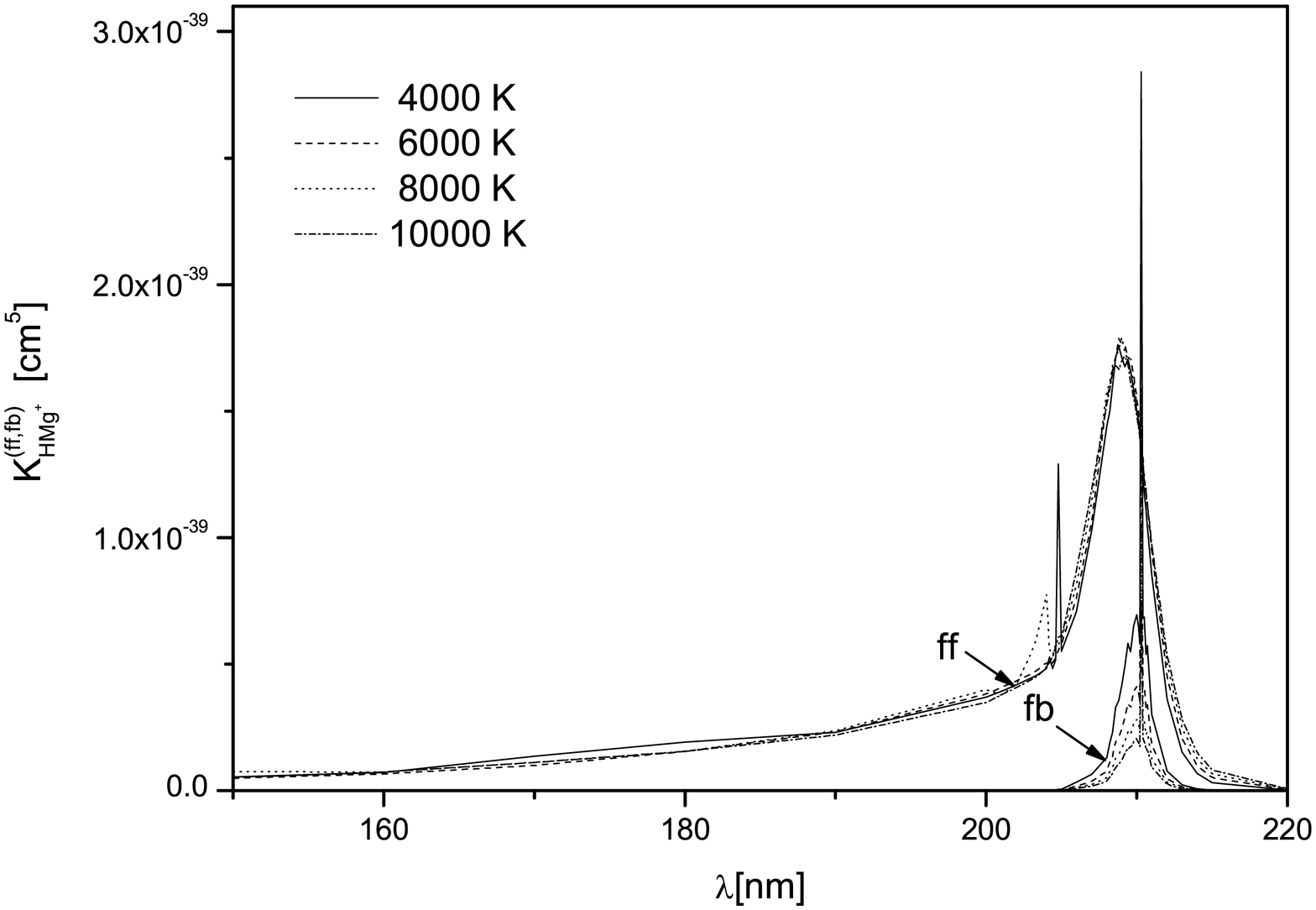} \caption{Same as in Fig.\ref{fig:HeH_Kfffb}, but
for HMg$^{+}$.}
\label{fig:HMg_Kfffb}
\end{center}
\end{figure}
%%%%%%%%%%%%%%%%%%%%%%%%%%%%%
\begin{figure}
\begin{center}
\includegraphics[width=\columnwidth,
height=0.75\columnwidth]{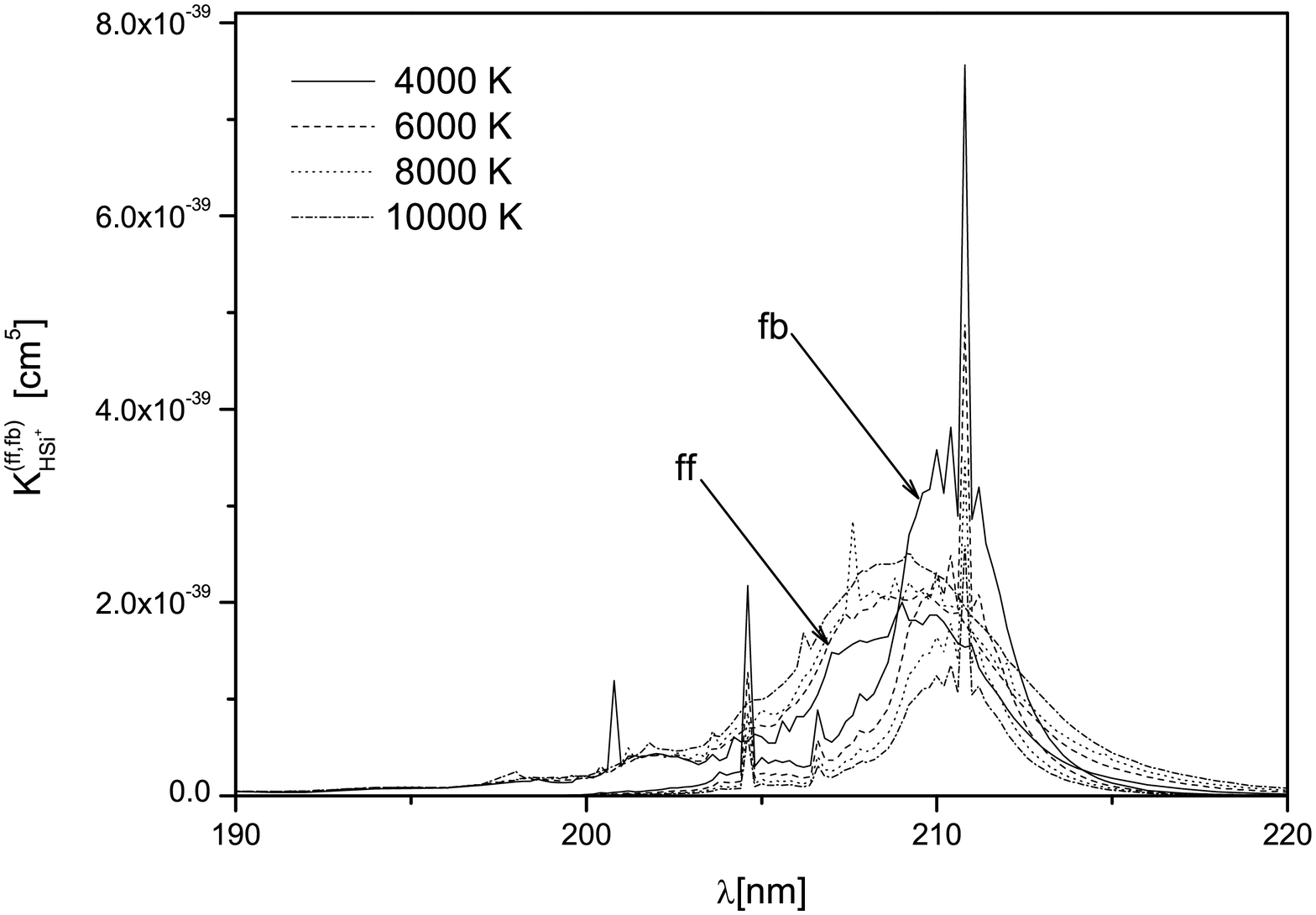} \caption{Same as in Fig.\ref{fig:HeH_Kfffb}, but
for the transition $X^{1}\Sigma^{+}\rightarrow B^{1}\Sigma^{+}$ for HSi$^{+}$.}
\label{fig:HSi_Kfffb}
\end{center}
\end{figure}
%%%%%%%%%%%%%%%%%%%%%%%%%%%%%%%%%%%%%%%%%%%%
\section{Results and Discussion}

\begin{figure}
\begin{center}
\includegraphics[width=\columnwidth,
height=0.75\columnwidth]{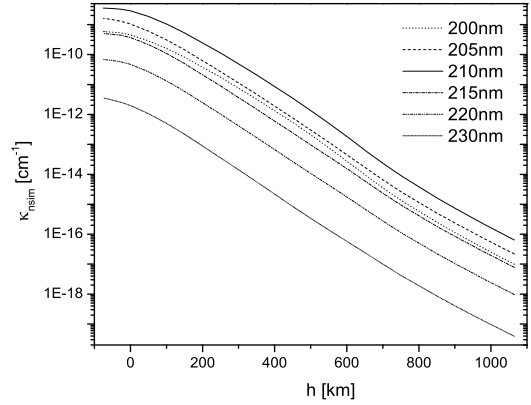}
\caption{Quiet Sun. Spectral absorption coefficient $\kappa_{nsim}(\lambda,T)$, given by
equation (\ref{eq:sumtot}) for 200 nm$ \le \lambda \le$ 230 nm.} \label{fig:200_230k}
\end{center}
\end{figure}

\begin{figure}
\begin{center}
\includegraphics[width=\columnwidth,
height=0.75\columnwidth]{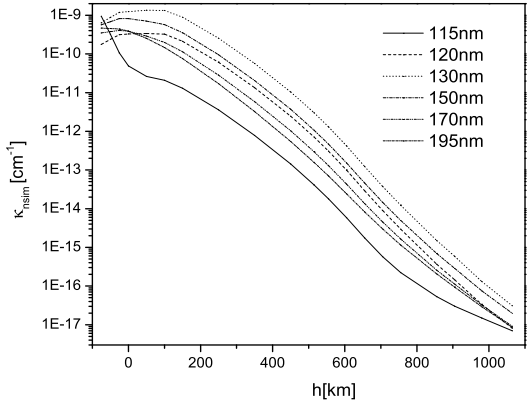}
\caption{Same as in Fig.\ref{fig:200_230k}, but for 115 nm$ \le \lambda \le$ 195 nm.}
\label{fig:120_155k}
\end{center}
\end{figure}

\begin{figure}
\begin{center}
\includegraphics[width=\columnwidth,
height=0.75\columnwidth]{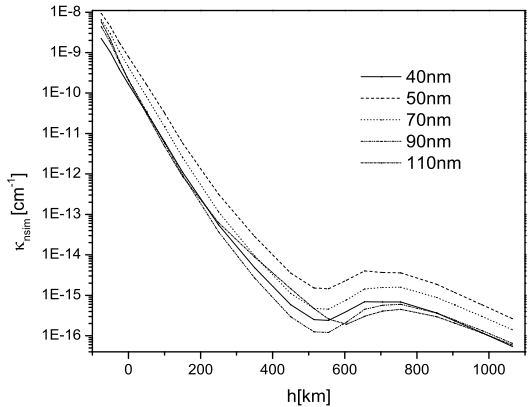}
\caption{Same as in Fig.\ref{fig:200_230k}, but for 40 nm$ \le \lambda \le$ 110 nm.}
\label{fig:40_80k}
\end{center}
\end{figure}

The absorption coefficients, as functions of the plasma's temperature and the atomic
and ionic densities in the solar photosphere, are determined here based on
the non-equilibrium model C from \citet{ver81}, where these parameters are presented
as functions of the height ($h$) of the considered layer with respect to the chosen
referent layer.
The total non-symmetric spectral absorption coefficient $\kappa_{ia;nsim}(\lambda)$ is
taken here, in accordance with Eq. (\ref{eq:kapatot}), in the form
%%%%%%%%%%%%%%%%%%%%%%%%%%%%%%%%%%%%%%%%%%%%%%%%%%%%%%%%%%%%%%%%%%
\begin{equation}
\label{eq:sumtot}
\kappa_{ia;nsim}(\lambda) = \kappa_{\textrm{HeH}^{+}}(\lambda) + \kappa_{\textrm{HMg}^{+}}(\lambda) +
\kappa_{\textrm{HSi}^{+}}(\lambda),
\end{equation}
%%%%%%%%%%%%%%%%%%%%%%%%%%%%%%%%%%%%%%%%%%%%%%%%%%%%%%%%%%%%%%%%%%
where the partial spectral absorption coefficients $\kappa_{AB^{+}}(\lambda)$ are determined
using the above expressions for the $bf$, $ff$ and $fb$ rate coefficients. The
results of the calculations of $\kappa_{ia;nsim}(\lambda)$ as a function of $h$, for -75
km$ \le h \le $1100 km\, are presented in Figs. \ref{fig:200_230k}-\ref{fig:40_80k} which
cover the part of UV and EUV region where 40 nm$ \le \lambda \le $230 nm. Consequently,
this part covers all regions of $\lambda$ relevant for the considered ion-atom systems
(see Figs. \ref{fig:HMg_sigphd}-\ref{fig:HSi_Kfffb}). In accordance with this, Fig. \ref{fig:200_230k}
illustrates the common contribution of the HMg$^{+}$ and HSi$^{+}$ absorption continua,
while Fig. \ref{fig:120_155k} refers to the region of exclusive domination of the HMg$^{+}$ continuum. Finally,
the Fig. \ref{fig:40_80k} illustrates the HeH$^{+}$ absorption continuum.
%%%%%% kapa %%%%
%%%%%%%%%%%%%%%%
%%%%%%%%%%%%%%%%
%%%%%%%%%%%%%%%%%%%%%%%%%%%%%%%%%%%%%%%%%%%%%%%

As the characteristics of the non-symmetric
absorption processes (\ref{eq:nonsim1})-(\ref{eq:nonsim3}), in the context of
their influence on the solar atmosphere opacity, here it is used the quantity
$G^{(nsim)}_{tot}(\lambda)$ defined by the relations
%%%%%%%%%%%%%%%%%%%%%%%%%%%%%%%%%%%%%%%%%%%%%%%%%%%%%%%%%%%%%%%%%%
\begin{equation}
\label{eq:Gnsim}
G^{(nsim)}_{tot}(\lambda) = \frac{\kappa_{ia;nsim}(\lambda)}{\kappa_{ia;tot}(\lambda)},
\kappa_{ia;tot}(\lambda) = \kappa_{ia;nsim}(\lambda) + \kappa_{ia;sim}(\lambda),
\end{equation}
%%%%%%%%%%%%%%%%%%%%%%%%%%%%%%%%%%%%%%%%%%%%%%%%%%%%%%%%%%%%%%%%%%
where $\kappa_{ia;sim}(\lambda)$ characterize the contribution of the
symmetric ion-atom absorption processes (\ref{eq:sim1}) and (\ref{eq:sim2}). In accordance with these relations
the quantity $G^{(nsim)}_{tot}(\lambda)$ describes the
relative contribution of the non-symmetric processes (\ref{eq:nonsim1})-(\ref{eq:nonsim3})
to the total absorption caused by all ion-atom absorption processes. Let us
note that the use of $\kappa_{ia;sim}(\lambda)$ as a referent quantity is justified
as these symmetric processes are now already included in some solar atmosphere models \citep{fon09}.

It is clear that apart of the quantity $G^{(nsim)}_{tot}(\lambda)$ as the characteristics of the processes (\ref{eq:nonsim1})-(\ref{eq:nonsim3}) could be used some other quantities, e.g. the ratio
$\kappa_{ia;tot}(\lambda)/\kappa_{ia;sim}(\lambda)$, which describes
the direct increase of the efficiency of the ion-atom absorption processes caused by the inclusion
of the non-symmetric ones. However, in the case of the solar atmosphere it would be very difficult to use this ratio. Namely,
in accordance with \citet{ver81} in the part of
the solar atmosphere around its temperature minimum the proton densities $N_{\textrm{H}^{+}} <<
N_{B+}$, where $B=$ Mg and Si, and consequently the quantity $[\kappa_{ia;nsim}(\lambda) +
\kappa_{ia;sim}(\lambda)]/\kappa_{ia;nsim}(\lambda) >> 1$. Consequently, the behavior of this
quantity can be hardly shown in the whole region of $h$ in the same proportion. Because of
that, as the characteristic of the significance of the non-symmetric absorption processes
(\ref{eq:nonsim1})-(\ref{eq:nonsim3}) for the solar atmosphere in UV and EUV region, the
quantity $G^{(nsim)}_{tot}(\lambda)$ is used, since from its definition follows that always
$0 < G^{(nsim)}_{tot}(\lambda) < 1$. The values of $\kappa_{ia;sim}(\lambda)$, needed for
the $G^{(nsim)}_{tot}(\lambda)$ determination, are taken from \cite{mih07a}.

\begin{figure}
\begin{center}
\includegraphics[width=\columnwidth,
height=0.75\columnwidth]{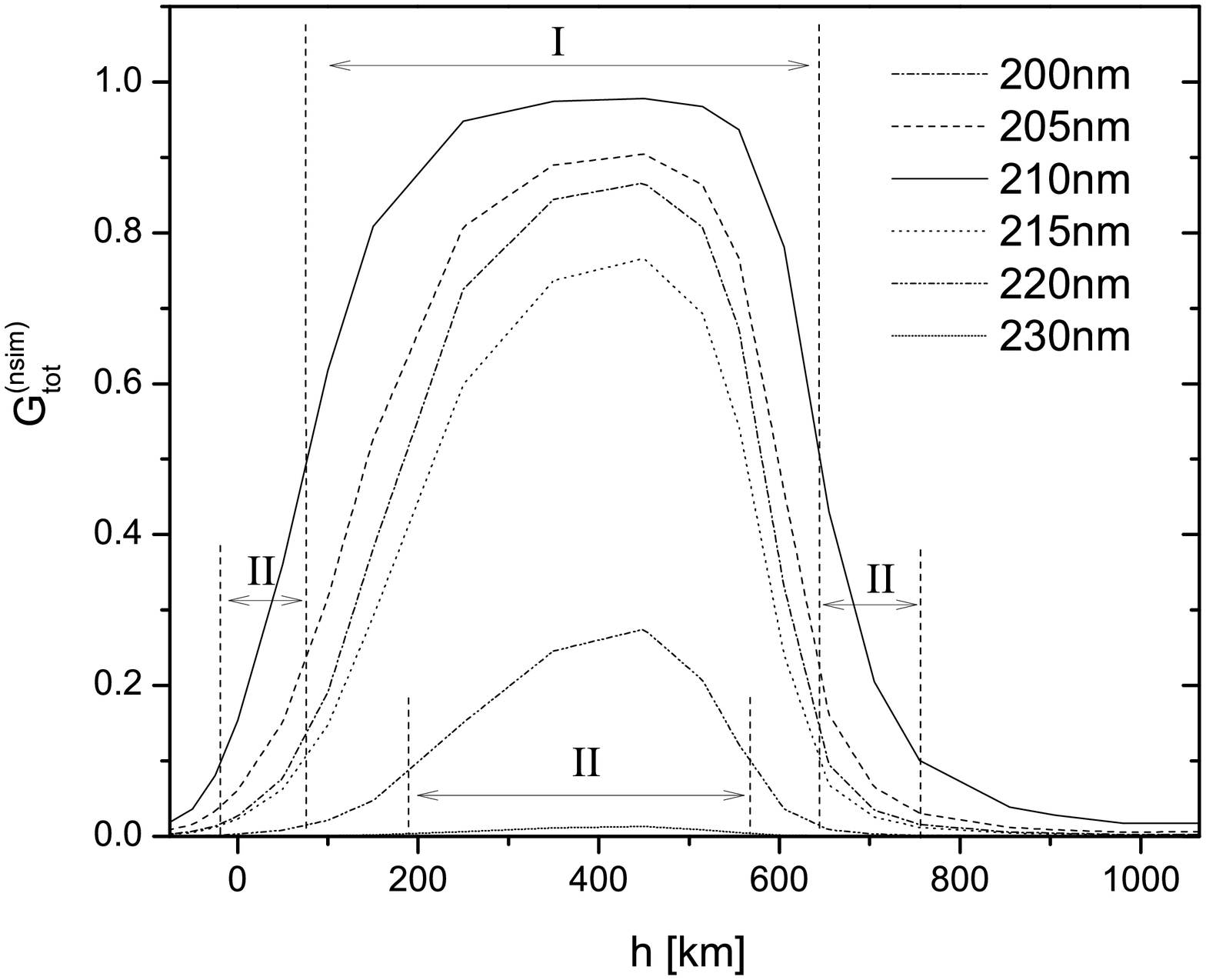}
\caption{The presented values of $G^{(nsim)}_{tot}(\lambda)$, given by equation (\ref{eq:Gnsim}),
as the function of $h$ for the quiet Sun for 200 nm$ \le \lambda \le$ 230 nm; I and II
are the regions of $h$ where $0.5 \lesssim G^{(nsim)}_{tot}(\lambda)$ and $0.1 \lesssim
G^{(nsim)}_{tot}(\lambda) < 0.5$ respectively.} \label{fig:200_230}
\end{center}
\end{figure}

\begin{figure}
\begin{center}
\includegraphics[width=\columnwidth,
height=0.75\columnwidth]{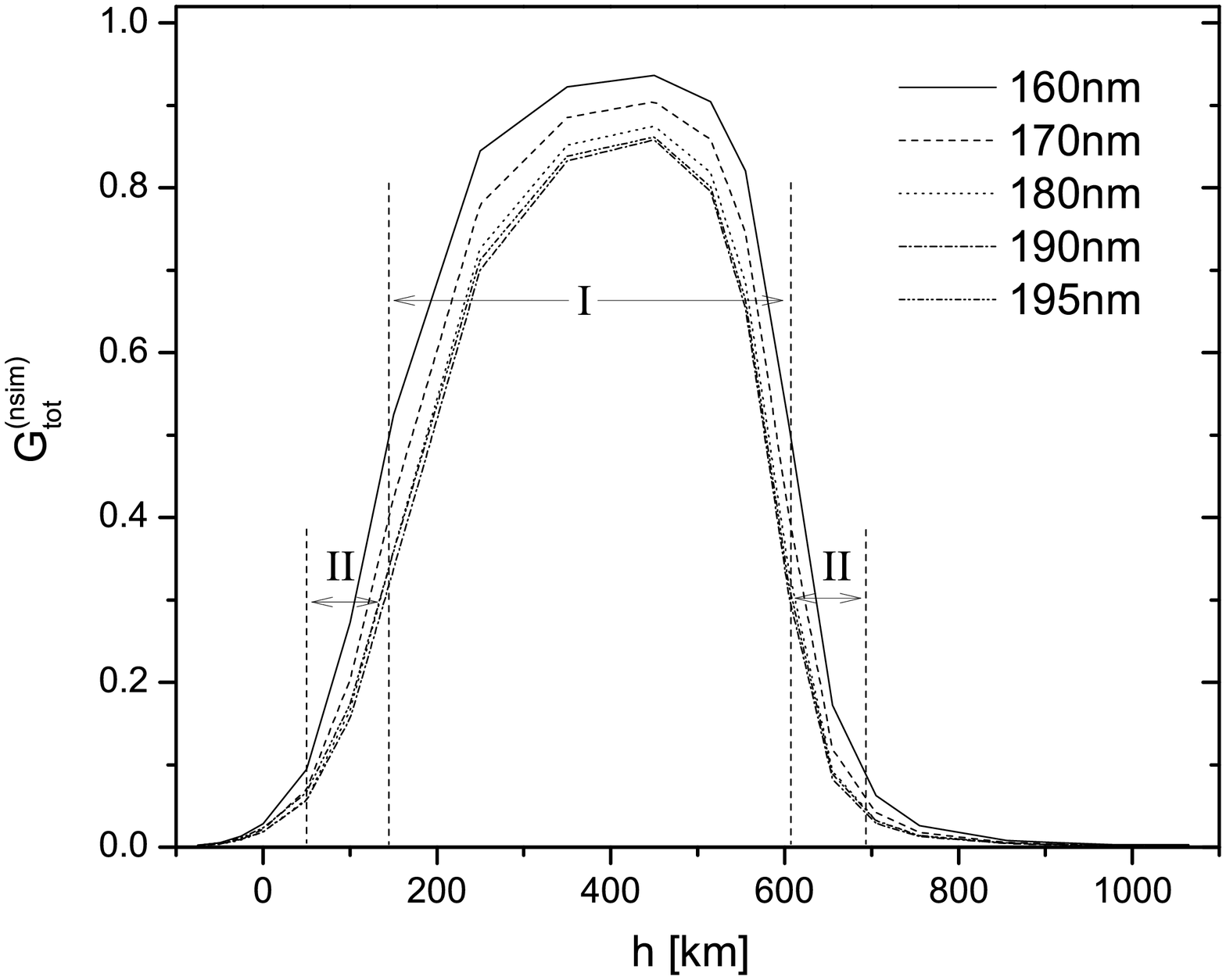}
\caption{Same as in Fig.\ref{fig:200_230}, but for 160 nm $\le \lambda \le$ 195 nm.}
\label{fig:160_195}
\end{center}
\end{figure}

\begin{figure}
\begin{center}
\includegraphics[width=\columnwidth,
height=0.75\columnwidth]{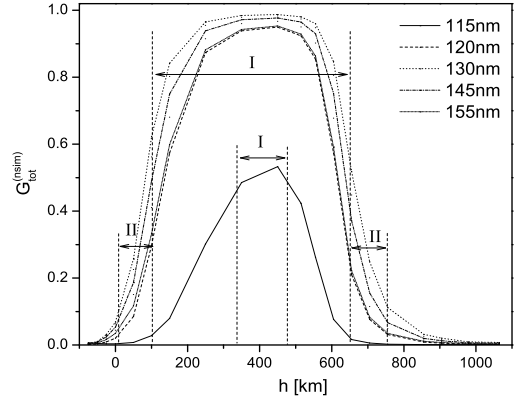}
\caption{Same as in Fig.\ref{fig:200_230}, but for 115 nm$ \le \lambda \le$ 155 nm.}
\label{fig:120_155}
\end{center}
\end{figure}

The calculated values of $G^{(nsim)}_{tot}(\lambda)$ as function of $h$, for the chosen
set of $\lambda$, are presented in Figs.\ref{fig:200_230}-\ref{fig:120_155}. From these
figures one can see that around the mentioned temperature minimum ($T \lesssim $5000
K,  150 km$ \lesssim h \lesssim $705 km) the contribution of non-symmetric processes
(\ref{eq:nonsim1})-(\ref{eq:nonsim3}) are dominant in respect to the symmetric
processes (\ref{eq:sim1}) and (\ref{eq:sim2}). Such region of the non-symmetric
processes domination is denoted in these figures as the region "I". Apart of that,
Figs. \ref{fig:200_230}-\ref{fig:120_155} show that within the rest of the considered region
of $h$ there are significant parts where the relative contribution of the non-symmetric
processes is close to or at least comparable with the contribution of the symmetric ones. In
the same figures these parts are denoted as regions "II".
%%%%%%% odnosi kapa %%%%%%%%%
%%%%%%%%%%%%%%%%%%%%%%%
%%%%%%%%%%%%%%%%%%%%%%%%%%%%%%%%%%%%%%%%%%%%
%%%%%%%%%%%%%%%%%%%%%%%%

\begin{figure}
\begin{center}
\includegraphics[width=\columnwidth,
height=0.75\columnwidth]{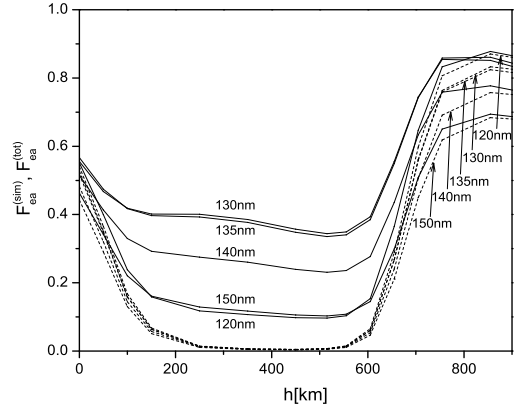} \caption{Quantities $F^{(sim)}_{ea}(\lambda)$
(dashed line) and $F^{(tot)}_{ea}(\lambda)$ (full line), defined in equations (\ref{eq:Fsim1}) and
(\ref{eq:Fsim2}) as the functions of $h$ for the Solar atmosphere for 120 nm $\le \lambda
\le$ 150 nm.}
\label{fig:H-120nm}
\end{center}
\end{figure}

In order to additionally show the importance of the considerations in the case of the solar
atmosphere of the non-symmetric processes (\ref{eq:nonsim1})-(\ref{eq:nonsim3}) here,
similarly to \citet{mih07a}, it was performed the comparison of the efficiencies of the
ion-atom absorption processes and the efficiency of such concurrent processes as the ion
H$^{-}$ photo-detachment and the electron-hydrogen atom inverse "bremsstrahlung" (H$^{-}$
continuum). Namely, among relevant concurrent absorption processes just these electron-atom
ones can be treated until now as the dominant in the spectral region which was considered
in \cite{mih07a}. For that purpose in this work it was compared the behavior of the quantity
%%%%%%%%%%%%%%%%%%%%%%%%%%%%%%%%%%%%%%%%%%%%%%%%%%%%%%%%%%%%%%%%%%
\begin{equation}
\label{eq:Fsim1}
F^{(sim)}_{ea}(\lambda) = \frac{\kappa_{ia;sim}(\lambda)}{\kappa_{ea}(\lambda)},
\end{equation}
%%%%%%%%%%%%%%%%%%%%%%%%%%%%%%%%%%%%%%%%%%%%%%%%%%%%%%%%%%%%%%%%%%
which is similar to the correspond quantity from \citet{mih07a} and characterize the relative
efficiency of the ion-atom symmetric processes and H$^{-}$ continuum, and the quantity
%%%%%%%%%%%%%%%%%%%%%%%%%%%%%%%%%%%%%%%%%%%%%%%%%%%%%%%%%%%%%%%%%%
\begin{equation}
\label{eq:Fsim2}
F^{(tot)}_{ea}(\lambda) = \frac{\kappa_{ia;tot}(\lambda)}{\kappa_{ea}(\lambda)},
\end{equation}
%%%%%%%%%%%%%%%%%%%%%%%%%%%%%%%%%%%%%%%%%%%%%%%%%%%%%%%%%%%%%%%%%%
which characterize the increasing of the total efficiency of the ion-atom radiative
processes after the including in the consideration of the non-symmetric processes
(\ref{eq:nonsim1})-(\ref{eq:nonsim3}). In these expressions $\kappa_{ia;tot}(\lambda)$ is
given by Eq.(\ref{eq:Gnsim}), the spectral absorption coefficient $\kappa_{ia;sim}(\lambda)$
characterizes the ion-atom symmetric processes (\ref{eq:sim1}) and (\ref{eq:sim2}) and is
taken from \citet{mih07a}, and the spectral absorption coefficient $\kappa_{ea}(\lambda)$
describes the H$^{-}$ continuum. In the case of the solar atmosphere $\kappa_{ea}(\lambda)$
is determined based on of \cite{sti70}, \cite{wis79} and \cite{ver81}. The behavior
of $F^{(sim)}_{ea}(\lambda)$ and $F^{(tot)}_{ea}(\lambda)$, as the functions of $h$, is
presented in Fig.\ref{fig:H-120nm}. This figure shows that the inclusion in the consideration
of the non-symmetric processes (\ref{eq:nonsim1})-(\ref{eq:nonsim3}) causes the significant
increases of the total efficiency of the ion-atom absorption processes, particulary in
the neighborhood of the solar atmosphere temperature minimum, where it become close to the
efficiency of the H$^{-}$ continuum.

In connection with the above mentioned let us note that according to \citet{ver81} in the
neighborhood of the solar atmosphere temperature minimum (50 km$ \lesssim h \lesssim $650
km) the $\textrm{Fe}$ component gives the maximal individual contribution to the electron density in
respect to the $\textrm{Mg}$ and $\textrm{Si}$ components. It means that the inclusion in the consideration
of the processes (\ref{eq:nonsim1})-(\ref{eq:nonsim3}) with $A=$ H and $B=$ Fe would surely
significantly increase the total contribution of the non-symmetric ion-atom absorption
processes, perhaps for about $50 \%$. Because of that we have as the task for the nearest
future to find the data about the relevant characteristics of the molecular ion HFe$^{+}$,
since the data from \citet{ver81} make possible to perform all needed calculations.

\begin{figure}
\begin{center}
\includegraphics[width=\columnwidth,
height=0.75\columnwidth]{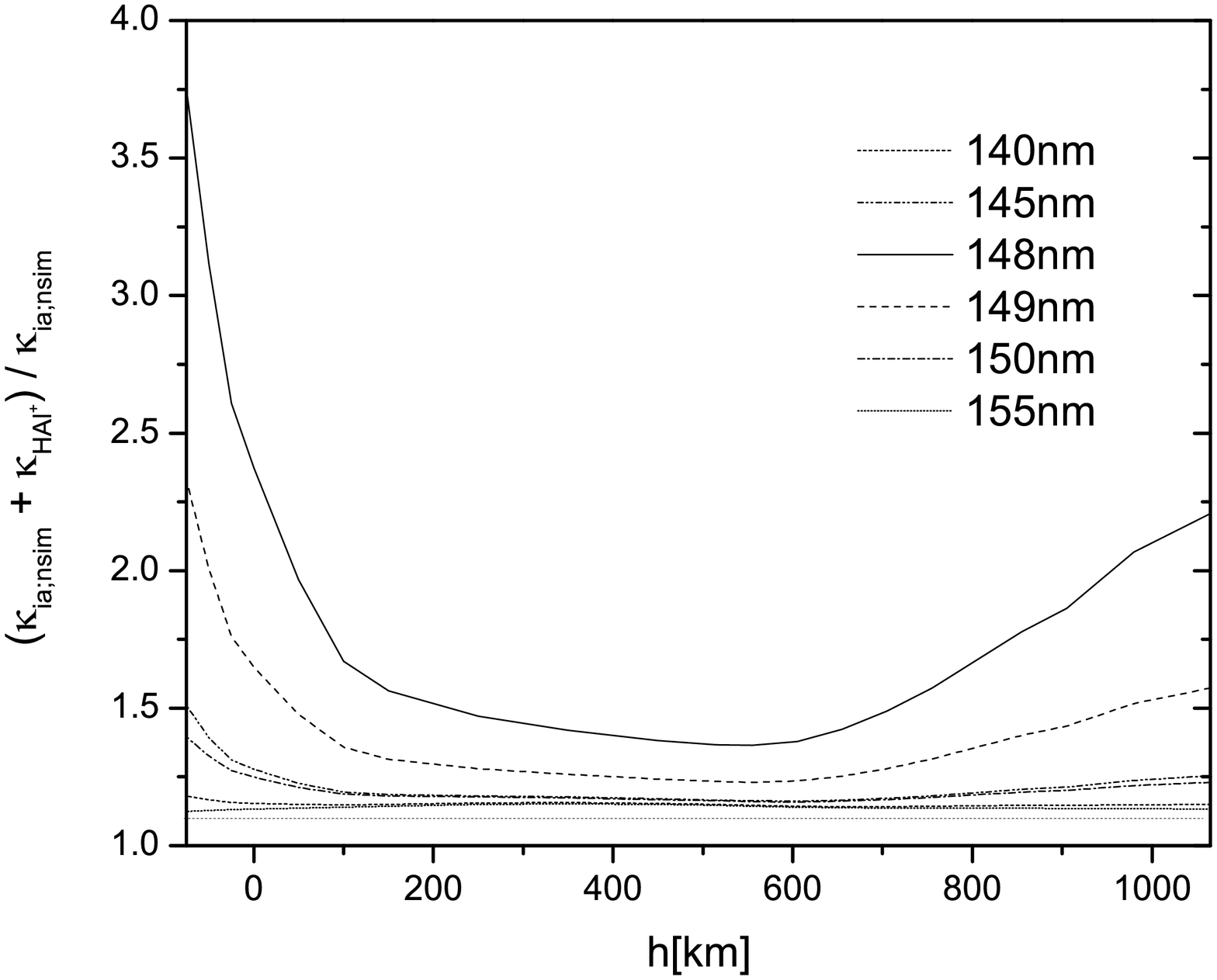} \caption{The ratio
$(\kappa_{ia;nsim}(\lambda)+\kappa_{AlH^{+}}(\lambda))/\kappa_{ia;nsim}(\lambda)$, where
$\kappa_{ia;nsim}(\lambda)$ is given by equation (\ref{eq:sumtot}) and $\kappa_{\textrm{AlH}^{+}}(\lambda)$
by equation (\ref{eq:kapa}) for $A=$ H and $B=$ Al, as a function of $h$ for the Solar atmosphere for
140 nm$ \le \lambda \le$ 155 nm.}
\label{fig:Alodnos}
\end{center}
\end{figure}

Let us note also that, according to the data from \citet{ver81} and \cite{fon09}, in the
solar atmosphere it should be include in the consideration the non-symmetric processes
(\ref{eq:nonsim1})-(\ref{eq:nonsim3}) with $A=$ H, where $B$ is the atom of the one of
such components ($\textrm{C}$, $\textrm{Al}$ etc.), which give visible contribution in narrow regions
of $h$ and $\lambda$. Now we have only the data needed for the case $A=$ H and $B=$ Al,
whose contribution is noticeable in the region 140 nm$ \lesssim \lambda \lesssim $155
nm. Since there is a certain difference between shapes of the lower potential curves
for the ion AlH$^{+}$, presented in Fig. \ref{fig:HAl_term} and the corresponding figure
from \cite{gue81}, whose nature by now is not completely clear, the contribution of the
processes (\ref{eq:nonsim1})-(\ref{eq:nonsim3}) with $A=$ H and $B=$ Al was not included
in the calculations described above. However, we think that this contribution can serve
to estimate the usefulness  of the inclusion in the consideration of these processes. For
that purpose we presented in Fig. \ref{fig:Alodnos} the results of the calculations of the
ratios $(\kappa_{ia;nsim}(\lambda)+\kappa_{\textrm{AlH}^{+}}(\lambda))/\kappa_{ia;nsim}(\lambda)$,
as function of $h$, for 140 nm$ \le \lambda \le $155 nm, where $\kappa_{ia;nsim}(\lambda)$
was determined according to Eq.\ref{eq:sumtot}. The behavior of this ratios presented in
above mentioned figure, in the region 200km $<h<$ 700km, is caused by the fast decrees of the
corresponding ion species concentration. This figure clearly demonstrate the fact that in the
significant parts of the solar photosphere the inclusion in the consideration of the processes
(\ref{eq:nonsim1})-(\ref{eq:nonsim3}) with $A=$ H and $B=$ Al should noticeably increase the
total contribution of the non-symmetric ion-atom processes. The above mentioned suggest that
the contribution of all metal components which by now were not included in the consideration
could significant increase the total efficiency of the ion-atom radiative processes, which
is characterized by the quantity $F^{(tot)}_{ea}(\lambda)$ in Fig.\ref{fig:H-120nm}.
%%%%%%%%%%%%%%%%%%%%%%%%%%%%%%%%%%%%%%%%%%%%%%%%
%%%%%%%%%%%%%%%%%%%%%%%%%%%%%%%%%%%%%%%%%%%%%%%%%5
%%%%% H- %%%%%%
%%%%%%%%%%%%%%%%%
The results presented in Figs.\ref{fig:200_230}-\ref{fig:120_155} and \ref{fig:H-120nm}
shows that the neglecting of the contribution of the non-symmetric processes
(\ref{eq:nonsim1})-(\ref{eq:nonsim3}) to the opacity of the solar atmosphere, in respect to
the contribution of symmetric processes (\ref{eq:sim1}) and (\ref{eq:sim2}) would caused
significant errors. From here it follows that the non-symmetric absorption processes
(\ref{eq:nonsim1})-(\ref{eq:nonsim3}) should be \emph{ab initio} included in the solar
atmosphere models.

\section{Conclusions}

From the presented material it follows that the considered non-symmetric ion-atom
absorption processes can not be treated only as one channel among many
equal channels with influence on the opacity of the solar atmosphere.
Namely, these non-symmetric processes around the temperature minimum increase
the absorption of the EM radiation, which is caused by all (symmetric and
non-symmetric) ion-atom absorption processes, so that this absorption becomes
almost uniform in the whole solar photosphere.
Moreover, the presented results
show that further investigations of these processes promise to demonstrate that
they are of similar importance as the known process of photo-detachment
of the ion H$^{-}$, which was treated until recently as absolutely dominant.
%%%%% sa i bez Al %%%%%%%%%%%%%%%%
%%%%%%%%%%%%%%%%%%%%%%%%%%%%%%%%%%%%%%%%%%%
Namely, the inclusion of the non-symmetric absorption processes into
consideration with $A=$ H and $B=$ Fe, as well as some other similar processes
(with $A=$ H and $B=$ Al etc), would significantly increase the contribution of
such processes to the solar-atmosphere opacity. All mentioned facts suggest that
the considered non-symmetric ion-atom absorption processes should be included
\emph{ab initio} in the solar-atmosphere models, as well as in the models of solar and near solar type stars.

\section*{Acknowledgments}

The authors wish to thank to Profs. V.N. Obridko and A.A. Nusinov
for the shown attention to this work. Also, the authors are thankful to the
Ministry of Education, Science and Technological Development of the Republic of
Serbia for the support of this work within the projects 176002, III4402.

%\bibliographystyle{mn2e.bst}
%\bibliography{B.bib}

\newcommand{\noopsort}[1]{} \newcommand{\printfirst}[2]{#1}
  \newcommand{\singleletter}[1]{#1} \newcommand{\switchargs}[2]{#2#1}

\label{lastpage}

\end{document}